\documentclass[a4paper,12pt]{article}

\usepackage{psfrag}
\usepackage{mathrsfs}

\usepackage{graphicx}
\usepackage[margin=15pt,font=small,labelfont=bf,labelsep=period]{caption}
\usepackage{amsmath}
\usepackage{amssymb}
\usepackage{amsfonts}
\usepackage{mathtools}
\usepackage{bm}
\usepackage{dsfont}
\usepackage{cite}
\usepackage{setspace}
\usepackage{environ,letltxmacro}
\usepackage{tikz}
\usetikzlibrary{arrows,decorations.pathreplacing}
\usepackage{enumitem}
\usepackage{hyperref}
\usepackage{relsize}
\usepackage[a4paper,textwidth=16.04cm,textheight=22cm,footskip=15mm]{geometry}

\usepackage{empheq}
\usepackage{environ}
\NewEnviron{boxalign}{\begin{empheq}[box=\fbox]{align} \BODY \end{empheq}}

\numberwithin{equation}{section}

\LetLtxMacro\oldequation\equation
\LetLtxMacro\endoldequation\endequation
\let\equation\relax
\let\endequation\relax
\NewEnviron{equation}[1][\empty]{%
 \ifx \empty#1
  \oldequation \BODY \endoldequation
 \else
  \ifx b#1
   \oldequation \fbox{$\displaystyle \BODY $} \endoldequation
  \else
   \oldequation \BODY \endoldequation
  \fi
 \fi}

\begin{document}

\begin{titlepage}

 \title{
Note on the super-extended Moyal formalism and its BBGKY hierarchy\\[2mm]
}

\date{}

\author{Carlo Pagani \\ [3mm]
{\small Institute of Physics, PRISMA \& MITP,}\\[-0.2em]
{\small Johannes Gutenberg University Mainz,}\\[-0.2em]
{\small Staudingerweg 7, D--55099 Mainz, Germany}
}

\maketitle
\thispagestyle{empty}

\vspace{2mm}
\begin{abstract}

We consider the path integral associated to the Moyal formalism for quantum mechanics extended
to contain higher differential forms by means of Grassmann odd fields. After revisiting
some properties of the functional integral associated to the (super-extended)
Moyal formalism, we give a convenient functional derivation of the
BBGKY hierarchy in this framework. In this case the distribution functions
depend also on the Grassmann odd fields.

\end{abstract}

\end{titlepage}

\newpage

\begin{spacing}{1.1}


\section{Introduction}

Recently a lot of work has been devoted to study at a deeper level
the formalism and the symmetries associated to the functional integral
of the Schwinger-Keldysh (SK) formalism \cite{Schwinger:1960qe,Keldysh:1964ud}, which allows to
deal with out of equilibrium phenomena in quantum field theory. It
turns out that a super-extension of the SK formalism has a rich symmetry structure
that allows to keep some feautures of the formalism manifest 
\cite{Crossley:2015evo,Glorioso:2016gsa,Gao:2017bqf,Glorioso:2017fpd,Haehl:2015foa,Haehl:2015uoc,Haehl:2016pec,Haehl:2016uah,Haehl:2017zac,Jensen:2017kzi}.

However, the Schwinger-Keldysh formalism is not the only possibility to
deal with a time dependent density matrix. In particular, Moyal introduced a formalism which naturally evolves the
density matrix and that allows to deal with quantum mechanics without
considering operators but only $c$-functions~\cite{Moyal:1949sk}. In this work we shall
revist the functional formulation of the super-extended Moyal formalism,
that allowed to put forward geometric notions, such as exterior derivatives,
in quantum mechanics, from the point of view of statistical mechanics.
Besides its own interest, this approach is tightly connected with
the recently proposed super-extension of the SK formalism. The main
objective of this work is to discuss the properties of the super-extended
Moyal formalism from the point of view of statistical mechanics and
give a functional derivation of the super-extended Bogoliubov-Born-Green-Kirkwood-Yvon (BBGKY) hierarchy
of distribution functions.

We shall review the Moyal approach to quantum mechanics in
section \ref{sec:Review-of-the-WM-formalism}, for the time being
it suffices to say that a special role is reserved to the density
matrix, whose time dependence is given in terms of the quantum analogue
of the Liouville operator, which, in the classical case, describes
the time evolution of the classical phase-space density distribution. 
The Moyal appraoch to quantum mechanics has also the nice
feature of displaying the classical limit in a very direct manner
since all the standard quantum mechanical operations, like the commutator
of two operators, are mapped to functions which are expressed as a
series expansion in $\hbar$.

The path integral associated to the Moyal formalism has been built
in~\cite{Marinov-PI,Gozzi:1993nk}. In~\cite{Gozzi:1993nk} the action appearing in the path
integral has been extended in order to contain new Grassmann odd fields,
which we will call ghosts, that allowed to put forward a proposal
for differential calculus in quantum mechanics. We will revisit this
path integral and briefly discuss its relation with the SK formalism
in section \ref{sec:Schwinger-Keldysh-formalism-and-ph-space-PI}.

In section \ref{sec:Super-extended-Moyal-formalism} we review the
super-extension of the Moyal formalism and its classical limit. In
doing so we discuss some properties of the associated functional integral.
In particular, we show that the introduction of the ghosts does not
modify the theory (i.e.~the correlation functions), consider the
associated topological properties and the relation with the super-extended
SK formalism.

Finally, in section \ref{sec:BBGKY-hierarchy} we give a functional derivation of the super-extended BBGKY hierarchy. 
In particular, we shall derive the BBGKY hierarchy via
a functional integral associated to the Moyal approach to quantum mechanics. 
More precisely, in section \ref{sub:Bosonic-BBGKY-hierarchy} we first discuss the case of the
path integral associated to the Moyal formalism and in section \ref{sub:Super-BBGKY-hierarachy}
we consider the extension containing the ghosts fields.

In section \ref{sec:summary-outlook} we summarize our findings and comment possible outlooks.

\section{Review of the Moyal approach to quantum mechanics and its
associated path integral \label{sec:Review-of-the-WM-formalism}}

\subsection{Moyal approach to quantum mechanics}

In this section we review the approach to quantum mechanics (QM) put
forward by Moyal~\cite{Moyal:1949sk}, building on previous work~\cite{Weyl,Wigner}
by Weyl and Wigner (see also~\cite{Groenwald,VanHove}).
The basic idea underlying Moyal formalism is that, ultimately, one
can formulate QM without using operators, rather only $c$-functions
are present and they are multiplied via the so called star-product
that will be introduced shortly. 
(We employ the notation used in~\cite{Gozzi:1993nk}.)

It is useful, however, to keep track of the relation between the standard
formalism and the Moyal one by considering the so called
``symbol calculus''. In particular, one sets up a one-to-one map
between the operators $\widehat{O}$ acting on a Hilbert space ${\cal V}$
and (complex valued) functions $O$ on a suitable manifold ${\cal M}$.
The relation between the space of functions, indicated with ${\cal F}\left({\cal M}\right)$,
and the operators 
is given by the following relation: given an operator $\widehat{O}$,
the associated function is provided by the ``symbol map'' $O=\mbox{symb}\left(\widehat{O}\right)$.
Furthermore, the space of symbols ${\cal F}\left({\cal M}\right)$
is equipped with the star-product $*$, which implements the multiplication
of operators at the level of their representative in ${\cal F}\left({\cal M}\right)$:
\begin{eqnarray*}
\mbox{symb}\left(\widehat{O}_{1}\widehat{O}_{2}\right) & = & \mbox{symb}\left(\widehat{O}_{1}\right)*\mbox{symb}\left(\widehat{O}_{2}\right)\,.
\end{eqnarray*}
This product is associative but non-commutative, thus keeping
track of these crucial features of the operatorial formulation of QM.

We shall consider a particularly important realization of the ideas
that we have just outlined by employing the Weyl symbol~\cite{Weyl}.
Essentialy, Weyl introduced an association rule mapping $c$-function
in the phase-space to operators with a specific ordering prescription. Before
introducing explictly this correspondence, let us set up our notation
for the phase-space. We denote all the phase-space variables collectively as $\varphi^{a}\equiv\left(q_{1},\cdots q_{n},p_{1},\cdots,p_{n}\right)$.
The symplectic two-form $\omega_{ab}$ and its inverse $\omega^{ab}$
can be written locally as 
\begin{eqnarray*}
\omega_{ab} & \equiv & \left(\begin{array}{cc}
0 & -\mathbb{I}\\
\mathbb{I} & 0
\end{array}\right) \, , \\
\omega^{ab} & \equiv & \left(\begin{array}{cc}
0 & \mathbb{I}\\
-\mathbb{I} & 0
\end{array}\right)\,.
\end{eqnarray*}
The Hamilton equation of motion and the Poisson brackets can then
be written respectively as
\begin{eqnarray*}
\dot{\varphi}^{a} & = & \omega^{ab}\partial_{b}H \, , \\
\left\{ f,g\right\} _{pb} & = & \partial_{a}f\omega^{ab}\partial_{b}g\,=\,\partial_{q}f\partial_{p}g-\partial_{p}f\partial_{q}g\,,
\end{eqnarray*}
where the subscript $pb$ allows to distinguish the Poisson brackets
from the Moyal brackets that will be introduced later on.

The Weyl symbol $O=\mbox{symb}\left(\widehat{O}\right)$ is defined as 
\begin{eqnarray*}
O\left(\varphi^{a}\right) & \equiv & \int\frac{d^{2n}\varphi_{0}^{a}}{(2\pi\hbar)^{n}}\exp\left[\frac{i}{\hbar}\varphi_{0}^{a}\omega_{ab}\varphi^{a}\right]\mbox{Tr}\left[\widehat{T}\left(\varphi_{0}\right)\widehat{O}\right] \,, 
\end{eqnarray*}
with
\begin{eqnarray*}
\widehat{T} \left(\varphi_{0}\right)& \equiv & \exp\left[\frac{i}{\hbar}\varphi_{0}^{a}\omega_{ab}\widehat{\varphi}^{b}\right]=\exp\left[\frac{i}{\hbar}\left(p_{0}\widehat{q}-q_{0}\widehat{p}\right)\right]\,.
\end{eqnarray*}
The inverse map is given by
\begin{eqnarray*}
\widehat{O} & = & \int\frac{d^{2n}\varphi_{0}^{a}d^{2n}\varphi^{a}}{(2\pi\hbar)^{2n}}O\left(\varphi^{a}\right)\exp\left[\frac{i}{\hbar}\varphi^{a}\omega_{ab}\varphi_{0}^{b}\right]\widehat{T}\left(\varphi_{0}\right)\\
 & = & \int\frac{d^{2n}\varphi_{0}^{a}d^{2n}\varphi^{a}}{(2\pi\hbar)^{2n}}O\left(\varphi^{a}\right)\exp\left[\frac{i}{\hbar}\left(p_{0}\left(\widehat{q}-q\right)-q_{0}\left(\widehat{p}-p\right)\right)\right]\,.
\end{eqnarray*}
A particularly important case is that of the density matrix.
Considering a pure state density matrix $\widehat{\rho}=|\psi\rangle\langle\psi|$ 
we have
\begin{eqnarray*}
\rho\left(\varphi^{a}\right) & = & \int\frac{d^{2n}\varphi_{0}^{a}}{(2\pi\hbar)^{n}}\exp\left[\frac{i}{\hbar}\varphi_{0}^{a}\omega_{ab}\varphi^{a}\right]\mbox{Tr}\left[\widehat{T}\left(\varphi_{0}\right)|\psi\rangle\langle\psi|\right]\\
 & = & \int d^{n}q_{0} \,
 e^{-\frac{i}{\hbar}pq_{0}}\psi\left(q+\frac{q_{0}}{2}\right)\psi^{\dagger}\left(q-\frac{q_{0}}{2}\right)\,.
\end{eqnarray*}
The so defined $\rho\left(q,p\right)$ is called the Wigner function
(WF). While the WF does not have the meaning of a probability density as
it can be also negative, standard quantum probability densities are
retrieved via the following integrations:
\begin{eqnarray*}
\left|\psi\left(q\right)\right|^{2} & = & \int\frac{d^{n}p}{\left(2\pi\hbar\right)^{n}}\rho\left(p,q\right) \, , \\
\left|\widetilde{\psi}\left(p\right)\right|^{2} & = & \int\frac{d^{n}q}{\left(2\pi\hbar\right)^{n}}\rho\left(p,q\right)\,.
\end{eqnarray*}
The expectation value of an observable is given by
\begin{eqnarray*}
\langle\psi|\widehat{O}|\psi\rangle & = & \int\frac{d^{n}qd^{n}p}{\left(2\pi\hbar\right)^{n}}\rho\left(p,q\right)O\left(p,q\right)\,.
\end{eqnarray*}

By construction, it is guaranteed that this formalism is {\it fully equivalent}
to QM. Indeed, by introducing the star-product we have an isomorphism
between operators and their multiplication, and the phase-space functions.
Consider the operator $\widehat{O}_{3}=\widehat{O}_{1}\widehat{O}_{2}$,
its Weyl symbol $O_3$ satisfies
\begin{eqnarray*}
\widehat{O}_{3} & = & \int\frac{d^{2n}\varphi_{0}^{a}d^{2n}\varphi^{a}}{(2\pi\hbar)^{2n}}O_{3}\left(\varphi^{a}\right)\exp\left[\frac{i}{\hbar}\varphi^{a}\omega_{ab}\varphi_{0}^{b}\right]\widehat{T}\left(\varphi_{0}\right)\,,
\end{eqnarray*}
where 
\begin{eqnarray*}
O_{3}\left(\varphi^{a}\right) & \equiv & \left(O_{1}*O_{2}\right)\left(\varphi^{a}\right)\\
 & = & O_{1}\left(\varphi^{a}\right)\exp\left[i\frac{\hbar}{2}\overleftarrow{\partial}_{a}\omega^{ab}\overrightarrow{\partial}_{b}\right]O_{2}\left(\varphi^{a}\right)\,,
\end{eqnarray*}
where we introduced the so called $*$-product.

We can then define the Moyal brackets (indicated with the subscript
$mb$)~\cite{Moyal:1949sk}
\begin{eqnarray*}
\left\{ O_{1},O_{2}\right\} _{mb} & \equiv & \frac{1}{i\hbar}\left[O_{1}*O_{2}-O_{2}*O_{1}\right]\\
 & = & \mbox{symb}\left(\frac{1}{i\hbar}\left[\widehat{O}_{1},\widehat{O}_{2}\right]\right)\,.
\end{eqnarray*}
Interestingly, the $\hbar\rightarrow0$ limit of the Moyal brackets
gives back the standard Poisson brackets of classical mechanics:
\begin{eqnarray*}
\lim_{\hbar\rightarrow0}\left\{ O_{1},O_{2}\right\} _{mb} & = & \left\{ O_{1},O_{2}\right\} _{pb}\,.
\end{eqnarray*}
Thus, already at this stage, it is evident that the classical limit
is particularly clear in this formalism. 

Furthermore, we wish to consider the time evolution of the density
matrix, which is given by
\begin{eqnarray*}
\partial_{t}\widehat{\rho} & = & -\frac{1}{i\hbar}\left[\widehat{\rho},\widehat{H}\right]\,.
\end{eqnarray*}
By applying the symbol map we have the evolution equation of the WF
$\rho\left(q,p;t\right)$:
\begin{eqnarray*}
\partial_{t}\rho & = & -\left\{ \rho,H\right\} _{mb}\,.
\end{eqnarray*}
It is clear that in the classical limit $\hbar\rightarrow0$ we obtain
\begin{eqnarray*}
\partial_{t}\rho & = & -\left\{ \rho,H\right\} _{pb}\\
 & = & -\left(\partial_{a}\rho\omega^{ab}\partial_{b}H\right)\,=\,\omega^{ab}\partial_{a}H\partial_{b}\rho\\
 & \equiv & -\widehat{L}\rho\,,
\end{eqnarray*}
where $\widehat{L}=\omega^{ab}\partial_{b}H\partial_{a}$ is the Liouville
operator. On the other hand, the fully quantum evolution of the WF is given by:
\begin{eqnarray*}
\partial_{t}\rho & = & -\left\{ \rho,H\right\} _{mb}\\
 & = & -\frac{1}{i\hbar}\left(\exp\left[i\frac{\hbar}{2}\omega^{ab}\partial_{a}^{\left(\rho\right)}\partial_{b}^{\left(H\right)}\right]\rho H-\exp\left[i\frac{\hbar}{2}\omega^{ab}\partial_{a}^{\left(H\right)}\partial_{b}^{\left(\rho\right)}\right]\rho H\right)\,.
\end{eqnarray*}
Writing
\begin{eqnarray*}
\exp\left[i\frac{\hbar}{2}\omega^{ab}\partial_{a}^{\left(\rho\right)}\partial_{b}^{\left(H\right)}\right] & = & \sum_{n=0}^{\infty}\frac{1}{n!}\left(i\frac{\hbar}{2}\omega^{ab}\partial_{a}^{\left(\rho\right)}\partial_{b}^{\left(H\right)}\right)^{n}\,,
\end{eqnarray*}
we can re-express the evolution equation for the WF as follows:
\begin{eqnarray*}
\partial_{t}\rho & = & -\frac{2}{\hbar}\left[\sum_{n=0}^{\infty}\frac{\left(-1\right)^{n}}{\left(2n+1\right)!}\left(\frac{\hbar}{2}\omega^{ab}\partial_{b}^{\left(H\right)}\partial_{a}^{\left(\rho\right)}\right)^{2n+1}H\rho\right]\\
 & = & -\frac{2}{\hbar}\sin\left(\frac{\hbar}{2}\omega^{ab}\partial_{b}^{\left(H\right)}\partial_{a}^{\left(\rho\right)}\right) H \rho \,\equiv\,-\widehat{L}_{\hbar}\rho\,.
\end{eqnarray*}
Finally, one can write the (generic) Moyal brackets as
\begin{eqnarray*}
\left\{ A,B\right\} _{mb} & = & A\frac{2}{\hbar}\sin\left(\frac{\hbar}{2}\overleftarrow{\partial}_{a}\omega^{ab}\overrightarrow{\partial}_{b}\right)B\,.
\end{eqnarray*}
A generalization of the Moyal brackets including Grassmann odd variables can be introduced using the star-product between
Grassmann odd variables~\cite{Berezin:1980xw}, see~\cite{Gozzi:1993nk}.

\subsection{Path integral for the Moyal formalism \label{sub:Path-integral-for-WignerMoyal-formalism}}

In this section we present a simple derivation of the path integral
associated to the Moyal formalism following ref.~\cite{PI-for-pedestrians}. For a more detailed description
and more careful analysis we refer the reader to~\cite{Marinov-PI,Gozzi:1993nk}. 
Let us note that the Schr{\"o}dinger equation,
\begin{eqnarray}
i\hbar\partial_{t}\psi & = & \widehat{H}\psi\,,\label{eq:Schroedinger-equation}
\end{eqnarray}
is formally very similar to the evolution equation for the Wigner
function, which indeed can be written as:
\begin{eqnarray}
i\partial_{t}\rho & = & -i\,\frac{2}{\hbar}\sin\left(\frac{\hbar}{2}\omega^{ab}\partial_{b}^{\left(H\right)}\partial_{a}^{\left(\rho\right)}\right)H\rho\,=\,-i\widehat{L}_{\hbar}\rho\,,\label{eq:quantum-Liouville-equation}
\end{eqnarray}
where $-i\widehat{L}_{\hbar}$ is a sort of ``Hamiltonian'' operator
and the Planck constant in (\ref{eq:Schroedinger-equation})
takes the value $\hbar=1$ in the LHS of (\ref{eq:quantum-Liouville-equation}).
Moreover, in the present case the phase-space variables $\varphi^{a}$
play the role of the position operator $\hat{x}$ of standard QM.
Therefore we need to introduce a conjugate variable to $\varphi^{a}$,
such conjugate variable is indicated with $\lambda_{a}$. 
Now, invoking the QM commutators between conjugate variables, 
i.e.~$\left[\hat{x} ,\hat{p}\right] = i\hbar$, we define, in full analogy,
$ \left[\widehat{\varphi}^{a},\widehat{\lambda}_{b}\right] = i\delta_{b}^{a} $~\cite{Gozzi:1993nk}.

In the $\varphi$-representation, $\widehat{\varphi}^{a}={\varphi}^{a}$, 
the conjugate operator reads $\widehat{\lambda}_{a}=-i\frac{\partial}{\partial\varphi^{a}}=-i\partial_{a}$.
Thus, we can rewrite
\begin{eqnarray*}
i\partial_{t}\rho & = & -i\frac{2}{\hbar}H\sin\left(\frac{\hbar}{2}\omega^{ab}\overleftarrow{\partial}_{b}\left(\partial_{a}=i\widehat{\lambda}_{a}\right)\right)\rho\\
 & = & -i\frac{2}{\hbar}H\sin\left(i\frac{\hbar}{2}\omega^{ab}\overleftarrow{\partial}_{b}\widehat{\lambda}_{a}\right)\rho
 \, \equiv \, \widetilde{{\cal H}}_{B}^{\hbar} \rho \,,
\end{eqnarray*}
where we defined the new ``Hamiltonian'' $\widetilde{{\cal H}}_{B}^{\hbar}$.
As a function of $\varphi^{a}$ and $\lambda_{a}$ the Hamiltonian
$\widetilde{{\cal H}}_{B}^{\hbar}$ reads
\begin{eqnarray*}
\widetilde{{\cal H}}_{B}^{\hbar} & = & -i\frac{2}{\hbar}\sin\left(i\frac{\hbar}{2}\lambda_{a}\omega^{ab}\partial_{b}\right)H\\
 & = & \frac{2}{\hbar}\left[\sum_{n=0}^{\infty}\frac{1}{\left(2n+1\right)!}\left(\frac{\hbar}{2}\lambda_{a}\omega^{ab}\partial_{b}\right)^{2n+1}H\right]\,.
\end{eqnarray*}
It is straightfoward to check that $\widetilde{{\cal H}}_{B}^{\hbar}$
can be rewritten as
\begin{eqnarray*}
\widetilde{{\cal H}}_{B}^{\hbar} & = & \frac{1}{\hbar}\left[H\left(\varphi^{a}-\frac{\hbar}{2}\omega^{ab}\lambda_{b}\right)-H\left(\varphi^{a}+\frac{\hbar}{2}\omega^{ab}\lambda_{b}\right)\right]\,
\end{eqnarray*}
or via the following expression
\begin{eqnarray*}
\widetilde{{\cal H}}_{B}^{\hbar} & = & \frac{2}{\hbar}\sinh\left(\frac{\hbar}{2}\omega^{ab}\lambda_{a}\partial_{b}\right)H\,.
\end{eqnarray*}
For later purposes, let us check the classical limit of $\widetilde{{\cal H}}_{B}^{\hbar}$:
\begin{eqnarray*}
\widetilde{{\cal H}}_{B}^{\hbar\rightarrow0} & = & \lim_{\hbar\rightarrow0}\frac{1}{\hbar}\left[H\left(\varphi^{a}-\frac{\hbar}{2}\omega^{ab}\lambda_{b}\right)-H\left(\varphi^{a}+\frac{\hbar}{2}\omega^{ab}\lambda_{b}\right)\right]\\
 & = & -\partial_{a}H\omega^{ab}\lambda_{b}\,=\,\lambda_{a}\omega^{ab}\partial_{b}H\,.
\end{eqnarray*}

In QM we know that, given an Hamiltonian $H$, the associated path
integral weight is characterized by an action $S=\int dt\left[p\dot{q}-H\right]$.
Therefore, the action associated to $\widetilde{{\cal H}}_{B}^{\hbar}$
is given by
\begin{eqnarray*}
S_{{\rm M}} & = & \int dt\left[\lambda_{a}\dot{\varphi}^{a}-\widetilde{{\cal H}}_{B}^{\hbar}\right]\,,
\end{eqnarray*}
where the subscript M reminds of Moyal. In the classical limit, we
obtain the action 
\begin{eqnarray*}
S_{\hbar\rightarrow0}^{{\rm M}} & \stackrel{\hbar\rightarrow0}{=} & \int dt\left[\lambda_{a}\dot{\varphi}^{a}-\lambda_{a}\omega^{ab}\partial_{b}H\right]\,,
\end{eqnarray*}
which is indeed known to be the action for the path integral formulation
of classical mechanics~\cite{Gozzi:1989bf}. We will study in more detail the so called classical path integral
(CPI) in section \ref{sub:Classical-path-integral}.

For later purposes, let us come back to the Hamitonian $\widetilde{{\cal H}}_{B}^{\hbar}$
and write explicitly the combinations $\varphi^{a}-\frac{\hbar}{2}\omega^{ab}\lambda_{b}$
and $\varphi^{a}+\frac{\hbar}{2}\omega^{ab}\lambda_{b}$ in the case
of a single particle, i.e.~$\varphi^{a}=q,p$ and $\lambda_{a}=\lambda_{q},\lambda_{p}$.
We obtain
\begin{eqnarray*}
\widetilde{{\cal H}}_{B}^{\hbar} & = & \frac{1}{\hbar}\left[H\left(\varphi^{a}-\frac{\hbar}{2}\omega^{ab}\lambda_{b}\right)-H\left(\varphi^{a}+\frac{\hbar}{2}\omega^{ab}\lambda_{b}\right)\right]\\
 & = & \frac{1}{\hbar}\left[H\left(p+\frac{\hbar}{2}\lambda_{q},q-\frac{\hbar}{2}\lambda_{p}\right)-H\left(p-\frac{\hbar}{2}\lambda_{q},q+\frac{\hbar}{2}\lambda_{p}\right)\right]\,.
\end{eqnarray*}

\subsection{Comparison with the Schwinger-Keldysh formalism and its phase-space
path integral \label{sec:Schwinger-Keldysh-formalism-and-ph-space-PI}}

In this section we briefly review the SK formalism and make contact
with the formalism of section \ref{sub:Path-integral-for-WignerMoyal-formalism}.
For the sake of clarity we shall consider a simple one-dimensional
description. Given the density matrix $\rho_{0}$ at an initial time
$t_{0}$, the time evolution of the density matrix is given by
\begin{eqnarray*}
\widehat{\rho}\left(t\right) & = & U\left(t,t_{0}\right)\widehat{\rho}_{0}U^{\dagger}\left(t,t_{0}\right)\,.
\end{eqnarray*}
The matrix element $\langle x^{\prime}|\rho\left(t\right)|x\rangle$
can be written as
\begin{eqnarray*}
\langle x^{\prime}|\widehat{\rho}\left(t\right)|x\rangle & = & \langle x^{\prime}|U\left(t,t_{0}\right)\widehat{\rho}_{0}U^{\dagger}\left(t,t_{0}\right)|x\rangle\\
 & = & \int {dx_{0}dx_{0}^{\prime}\,} \langle x^{\prime}|U\left(t,t_{0}\right)|x_{0}\rangle\langle x_{0}|\widehat{\rho}_{0}|x_{0}^{\prime}\rangle\langle x_{0}^{\prime}|U^{\dagger}\left(t,t_{0}\right)|x\rangle\\
 & = & \int {dx_{0}dx_{0}^{\prime}\,}
 \langle x_{0}|\widehat{\rho}_{0}|x_{0}^{\prime}\rangle\int_{x_{1}\left(t_{0}\right)=x_{0}}^{x_{1}\left(t\right)=x^{\prime}}{\cal D}x_{1}e^{iS\left[x_{1}\right]}\int_{x_{2}\left(t_{0}\right)=x_{0}^{\prime}}^{x_{2}\left(t\right)=x}{\cal D}x_{2}e^{-iS\left[x_{2}\right]}\,.
\end{eqnarray*}
The trace is obtained via
\begin{eqnarray*}
\mbox{Tr}\left[\widehat{\rho}\left(t\right)\right] & = & \int {dx\,}\langle x|\widehat{\rho}\left(t\right)|x\rangle\\
 & = & \int {dx dx_{0}dx_{0}^{\prime}\,}
 \langle x_{0}|\widehat{\rho}_{0}|x_{0}^{\prime}\rangle\int_{x_{1}\left(t_{0}\right)=x_{0}}^{x_{1}\left(t\right)=x}\int_{x_{2}\left(t_{0}\right)=x_{0}^{\prime}}^{x_{2}\left(t\right)=x}{\cal D}x_{1}{\cal D}x_{2}e^{\frac{i}{\hbar}S\left[x_{1}\right]-\frac{i}{\hbar}S\left[x_{2}\right]}\,.
\end{eqnarray*}
The path integrals over $x_{1}$ and $x_{2}$ can be re-expressed
as path integrals in phase-space where one integrates also over the
momentum, which is not subject to any boundary condition. Ignoring
for a moment the factor $\langle x_{0}|\rho_{0}|x_{0}^{\prime}\rangle$,
which carries the information on the initial state, we rewrite
\begin{eqnarray*}
\int{\cal D}x_{1}{\cal D}x_{2}\,e^{\frac{i}{\hbar}S\left[x_{1}\right]-\frac{i}{\hbar}S\left[x_{2}\right]} & = & \int{\cal D}x_{1}{\cal D}x_{2}{\cal D}\pi_{1}{\cal D}\pi_{2}\,e^{\frac{i}{\hbar}\int\left(\pi_{1}\dot{x}_{1}-H\left(\pi_{1},x_{1}\right)\right)-\frac{i}{\hbar}\int\left(\pi_{2}\dot{x}_{2}-H\left(\pi_{2},x_{2}\right)\right)}\,.
\end{eqnarray*}
Furthermore, let us introduce the following symmetric and anti-symmetric combinations,
which we call $s$-fields and $a$-fields respectively:
\begin{eqnarray*}
x_{a} & \equiv & \frac{1}{\hbar}  \left(x_{1}-x_{2} \right) \\
x_{s} & \equiv & \frac{x_{1}+x_{2}}{2}\\
\pi_{a} & \equiv & \frac{1}{\hbar}  \left(\pi_{1}-\pi_{2} \right) \\
\pi_{s} & \equiv & \frac{\pi_{1}+\pi_{2}}{2}\,.
\end{eqnarray*}
Now, rewriting our path integral in terms of the new fields we have
\begin{eqnarray*}
\int{\cal D}x_{1}{\cal D}x_{2}\,e^{\frac{i}{\hbar}S\left[x_{1}\right]-\frac{i}{\hbar}S\left[x_{2}\right]} 
& = & 
\int{\cal D}x_{s}{\cal D}x_{a}{\cal D}\pi_{s}{\cal D}\pi_{a}\,
\exp\Biggr[i \int\Bigr(\pi_{s}\dot{x}_{a}+\pi_{a}\dot{x}_{s}+\\
 &  & \frac{1}{\hbar}H\left(\pi_{s}-\hbar \frac{\pi_{a}}{2},x_{s}-\hbar\frac{x_{a}}{2}\right)-H\left(\pi_{s}+\hbar\frac{\pi_{a}}{2},x_{s}+\hbar\frac{x_{a}}{2}\right)\Bigr)\Biggr]\,,
\end{eqnarray*}
so that final SK action reads
\begin{eqnarray*}
S_{{\rm SK}} & = & \int dt \left[\pi_{s}\dot{x}_{a}+\pi_{a}\dot{x}_{s}+\frac{1}{\hbar}H\left(\pi_{s}-\hbar\frac{\pi_{a}}{2},x_{s}-\hbar\frac{x_{a}}{2}\right)-\frac{1}{\hbar}H\left(\pi_{s}+\hbar\frac{\pi_{a}}{2},x_{s}+\hbar\frac{x_{a}}{2}\right)\right]\,.
\end{eqnarray*}
By performing an expansion in $\hbar$ we obtain, at the first non-trivial order, the following expression
\begin{eqnarray*}
S_{\hbar\rightarrow0}^{{\rm SK}} & = & \int dt 
\left[-x_{a}\Bigr(\dot{\pi}_{s}+\partial_{x_{s}}H\left(\pi_{s},x_{s}\right)\Bigr)
+\pi_{a}\Bigr(\dot{x}_{s}-\partial_{\pi_{s}}H\left(\pi_{s},x_{s}\right)\Bigr)\right]
+O\left(\hbar \right) \,.
\end{eqnarray*}
Interestingly enough, the terms isolated in the round brackets are
exactly the Hamilton equation of motion in terms of the $s$-fields. 
On the other hand, the $a$-fields appear only linearly  after the $\hbar$ expansion,
thus acquiring the status of Lagrange multipliers for the Hamilton's equation of motion.

However, not only the classical
limit coincides with its counterpart found in the path integral for the Moyal formalism. 
Actually the two path integrals are fully equivalent. To
see this explicitly let us first compare the classical limit of the actions
involved, i.e.~$S_{\hbar\rightarrow0}^{{\rm SK}}$ and $S_{\hbar\rightarrow0}^{{\rm M}}$.
We see that the two actions are the same provided that we make the following
identifications:
\begin{eqnarray*}
\pi_{s} & \equiv & p\\
x_{s} & \equiv & q\\
x_{a} & \equiv & -\lambda_{p}\\
\pi_{a} & \equiv & \lambda_{q}\,.
\end{eqnarray*}
If we implement this identification in the still exact $S_{{\rm SK}}$ we
obtain
\begin{eqnarray}
S_{{\rm SK}} & = & \int dt \left[\pi_{s}\dot{x}_{a}+\pi_{a}\dot{x}_{s}+\frac{1}{\hbar}H\left(\pi_{s}-\hbar\frac{\pi_{a}}{2},x_{s}-\hbar\frac{x_{a}}{2}\right)-\frac{1}{\hbar}H\left(\pi_{s}+\hbar\frac{\pi_{a}}{2},x_{s}+\hbar\frac{x_{a}}{2}\right)\right]\nonumber \\
 & = & \int dt \left[-p\dot{\lambda}_{p}+\lambda_{q}\dot{q}+\frac{1}{\hbar}H\left(p-\hbar\frac{\lambda_{q}}{2},q+\hbar\frac{\lambda_{p}}{2}\right)-\frac{1}{\hbar}H\left(p+\hbar\frac{\lambda_{q}}{2},q-\hbar\frac{\lambda_{p}}{2}\right)\right]\nonumber \\
 & = & \int dt \left[\lambda_{p}\dot{p}+\lambda_{q}\dot{q}-\widetilde{{\cal H}}_{B}^{\hbar}\right]+\mbox{surface term}\,.\label{eq:from-SK-SA-variables-to-Moyal}
\end{eqnarray}
This makes it clear that the functional integrals of the Moyal and Schwinger-Keldysh approach
are actually identical and are related by simple identifications:
the $s$-fields correspond to the Moyal's phase-space coordinates $\varphi=\left(q,p\right)$
while the $a$-fields play the role of the conjugate variables $\lambda =\left(\lambda_q,\lambda_p\right)$.

Let us note that the integration by parts performed in (\ref{eq:from-SK-SA-variables-to-Moyal})
leaves a surface term which is somewhat unpleasant. This surface term
is however not present after a more careful analysis. The kernel of
propagation associated to the SK action $S_{{\rm SK}}$ has the form
$\langle x_{1,t},x_{2,t};t|x_{1,0},x_{2,0};0\rangle$, where $x_{i,0}$
and $x_{i,t}$ are eigenvalues of the position operators $\widehat{x}_{1}$
and $\widehat{x}_{2}$. Equivalently, a state $|x_{1,t},x_{2,t};t\rangle$
can also be labelled as an eigenstate of the operators $\widehat{q}=\frac{1}{2}\left(\widehat{x}_{1}+\widehat{x}_{2}\right)$
and $\widehat{\lambda}_{p}=\left(\widehat{x}_{2}-\widehat{x}_{1} \right)/\hbar$ so that
we can rewrite $\langle x_{1,t},x_{2,t};t|x_{1,0},x_{2,0};0\rangle=\langle q_{t},\lambda_{p,t};t|q_{0},\lambda_{p,0};0\rangle$.
On the other hand, when we consider the path integral associated to
the Moyal action $S_{{\rm M}}$ we are actually considering the kernel
of propagation $\langle q_{t},p_{t};t|q_{0},p_{0};0\rangle$. The
two kernels are related as follows
\begin{eqnarray}
\langle q_{t},p_{t};t|q_{0},p_{0};0\rangle & = & \int d{\lambda_{p,t}} d{\lambda_{p,0}}\frac{1}{\left(\sqrt{2\pi}\right)^{2}}e^{i\lambda_{p,t}p_{t}-i\lambda_{p,0}p_{0}}\langle q_{t},\lambda_{p,t};t|q_{0},\lambda_{p,0};0\rangle\,.\label{eq:relation-Moyal-vs-SK-basis-for-kernels}
\end{eqnarray}
We observe that, using the path integral expression for the kernels
of propagation, the surface terms appearing in (\ref{eq:from-SK-SA-variables-to-Moyal})
cancel against the exponential appearing in (\ref{eq:relation-Moyal-vs-SK-basis-for-kernels}).

Before concluding this section let us point out some generic features
of the Schwinger-Keldysh formalism for later purposes. Let us define
the generating functional of connected Green's function via
\begin{eqnarray}
e^{W\left[J_{1};J_{2}\right]} & = & \mbox{Tr}\left[U_{1}\left(+\infty,-\infty;J_{1}\right)\rho_{0}U_{2}^{\dagger}\left(+\infty,-\infty;J_{2}\right)\right]\,,\label{eq:exp-W-J1-J2-via-trace}
\end{eqnarray}
where $J_{i}$ are sources which are associated to operator insertions
in the forward and backward evolutions. Clearly, if we set $J_{1}=J_{2}=J$
the cyclicity of the trace implies that (see for instance~\cite{Crossley:2015evo})
\begin{equation}
e^{W\left[J;J\right]}=1\,.\label{eq:unitarity-condition-SK}
\end{equation}
 This condition is sometimes referred to as the unitarity condition.

Moreover, if the initial density matrix is thermal, namely $\rho_{0}\sim e^{-\beta H}$,
then a further condition, known as the KMS condition, applies. Indeed, in this
case $\rho_{0}$ can be thought of as an evolution operator
for imaginary times. For an initial thermal density matrix, manipulating
the trace (\ref{eq:exp-W-J1-J2-via-trace}), one can connect $W\left[J_{1},J_{2}\right]$
to the generating functional associated to the time reversed process,
see e.g.~\cite{Crossley:2015evo,Sieberer:2015hba}. 
By combining the KMS condition with time reversal symmetry\footnote{
$PT$ and $CPT$ symmetries have also been used for the same purpouse
\cite{Crossley:2015evo,Jensen:2017kzi}.} 
one obtains a constraint directly on the functional $W\left[J_{1},J_{2}\right]$:
\begin{eqnarray}
W\left[J_{1}\left(t_{1}\right),J_{2}\left(t_{2}\right)\right] & = & W\left[J_{1}\left(-t_{1}\right),J_{2}\left(-t_{2}-i\beta\right)\right]\,.\label{eq:KMS-condition-on-W-for-cpt-invariant-case}
\end{eqnarray}
If the microscopic action $S$ is time reversal (and time translation)
invariant, it is possible to check that relation (\ref{eq:KMS-condition-on-W-for-cpt-invariant-case})
follows from a symmetry of the action in the forward and backward
branches~\cite{Crossley:2015evo,Gao:2017bqf,Jensen:2017kzi}.

\section{Super-extended Moyal formalism \label{sec:Super-extended-Moyal-formalism}}

In this section we consider the super-extended Moyal formalism put forward in~\cite{Gozzi:1993nk}
as a proposal to introduce an exterior differential calculus in QM.
This formalism naturally evolves not only the Wigner function $\rho \left( \varphi ;t \right)$ 
but also its generalization including additional Grassmann odd fields.

In section \ref{sub:Moyal-formalism} we show that 
the super-extended Moyal formalism yields the same correlation functions
as the standard one (provided one considers the very same observable).

In section \ref{sub:Comparison-with-the-SK-formalism} we compare
the super-extended Moyal formalism with the super-extended SK framework.
We pay particular attention to the topological sector of the theory
and show that by changing the boundary condition of the path integral
one can construct topological invariants of the phase-space manifold.

\subsection{Super-extended Moyal formalism \label{sub:Moyal-formalism}}

We have seen that the path integral description of the Moyal formalism
allows easily to make contact with the path integral description of
classical mechanics~\cite{Gozzi:1989bf}. It turns out that in this limiting
case a supersymmetric extension is very useful and it has been studied
thoroughly giving light to many interesting results related to ergodicity, the symplectic geometry
of classical mechanics, quantization, and so on~\cite{Gozzi:1989xz,Gozzi:1993tm,Gozzi:1999at,Abrikosov:2004cf,Gozzi:2009ip,Gozzi:2010iq,Cattaruzza:2010wc}.
In section \ref{sub:Classical-path-integral} we give a brief overview
of the CPI formalism (the reader already familiar with it may skip
this section). In section \ref{sub:Supersymmetric-extension-of-Moyal-formalism}
we review the super-extended Moyal formalism pointing out that the introduction of the ghosts
does not modify the correlation functions.

\subsubsection{Classical path integral \label{sub:Classical-path-integral}}

In the 1930s Koopman and von Neumann gave an operatorial
formulation of classical mechanics~\cite{koopman,vonNeumann}.
Let us consider the Liouville equation for a probability density in
the phase-space, $\rho(q,p;t)$:
\begin{eqnarray*}
\partial_{t}\rho\left(q,p\right) & = & \Bigr[\partial_{q}H\partial_{p}-\partial_{p}H\partial_{q}\Bigr]\rho\left(q,p;t\right)\,.
\end{eqnarray*}
Koopman and von Neumann, inspired by the formalism of quantum
mechanics, introduced an Hilbert space of the square integrable functions
on the phase-space and constructed an \textit{operatorial} approach
to CM~\cite{koopman,vonNeumann}. Given this operatorial formalism for CM,
we expect that there should be a path integral counterpart of it and,
of course, it should be different from the one associated to QM. This path integral approach
was studied in~\cite{Gozzi:1989bf}, whose main properties we now review.

First of all, we can consider that the transition amplitude for CM
should give weight ``one'' to the classical trajectories and zero
to all other paths. So the classical ``transition amplitude'' is:
\begin{eqnarray}
K\left(\varphi^{a},t;\varphi_{0}^{a},t_{0}\right) & = & {\delta}\left(\varphi^{a}-\Phi_{cl}^{a}(t;\varphi_{0}^{a},t_{0})\right)
\label{eq:CPI-initial-delta-function}
\end{eqnarray}
where $\Phi_{cl}^{a}(t;\varphi_{0}^{a},t_{0})$ is the classical solution
of the Hamiltonian equation of motion, 
\[
\dot{\varphi}^{a}=\omega^{ab}\frac{\partial H}{\partial\varphi^{b}}\,,
\]
with initial conditions $\Phi_{cl}^{a}(t_0;\varphi_{0}^{a},t_{0})=\varphi_{0}^{a}$. The functional Dirac delta
in (\ref{eq:CPI-initial-delta-function}) can be written as: 
\[
{\delta}\left(\varphi^{a}-\Phi_{cl}^{a}\right)=\delta\left(\dot{\varphi}-\omega^{ab}\partial_{b}H\right)\det\left(\delta_{b}^{a}\partial_{t}-\omega^{ac}\partial_{c}\partial_{b}H\right)\,,
\]
where $\mbox{det}\left[\cdots\right]$ is a functional determinant.
By expressing the Dirac delta as a Fourier transform and exponentiating
the determinant via Grassmann odd fields, to which we refer to as
ghosts, we finally obtain the representation:
\begin{eqnarray*}
K(\varphi^{a},t;\varphi_{0}^{a},t_{0}) & = & \int{\cal D}^{\prime\prime}\varphi^{a}{\cal D}\lambda_{a}{\cal D}c^{a}{\cal D}\bar{c}_{a}\exp\left[i\int_{t_{0}}^{t}d\tau\widetilde{{\cal L}}\right]\,,
\end{eqnarray*}
where the integration ${\cal D}^{\prime\prime}\varphi^{a}$ indicate
the integration over $\varphi^{a}$ with fixed initial and final configuration.
The Lagrangean $\widetilde{{\cal L}}$ which appears in the formula
is:
\begin{eqnarray}
\widetilde{{\cal L}} & = & {\displaystyle \lambda_{a}(\dot{\varphi}^{a}-\omega^{ab}\partial_{b}H)+i\bar{c}_{a}\dot{c}^{a}-i\bar{c}_{a}\omega^{ac}\partial_{c}\partial_{b}Hc^{b}}\label{eq:CPI-lagrangean}
\end{eqnarray}
and its associated Hamiltonian reads:
\begin{equation*}
\widetilde{{\cal H}}= \lambda_{a}\omega^{ab}\partial_{b}H+i\bar{c}_{a}\omega^{ac}\partial_{c}\partial_{b}Hc^{b}\,.
\end{equation*}

The action associated to $\widetilde{{\cal L}}$ enjoys a set of universal
symmetries. In particular, the action is invariant under the group
$ISp\left(2\right)$ and enjoys a $N=2$ supersymmetry~\cite{Gozzi:1989bf}.
The reader may wonder about the role of the ghosts $c^{a}$ and $\bar{c}_{a}$.
It turns out that these ghosts have been crucial to make contact with
many interesting topics like Cartan calculus, ergodicity and so on
(we refer the reader to~\cite{PI-for-pedestrians} for an overview). 

Before concluding, let us recall one of the universal symmetries present
in the CPI and consider the symmetry transformation
\begin{equation}
\delta\varphi^{a}=\epsilon c^{a},\;\;\delta\bar{c}_{a}=i\epsilon\lambda_{a},\;\;\delta c^{a}=\delta\lambda_{a}=0\,.\label{eq:BRS-symmetry-transf}
\end{equation}
Remarkably, the conserved charge associated to this symmetry, denoted
$Q_{{\rm BRS}}=ic^{a}\lambda_{a}$ in~\cite{Gozzi:1989bf}, can be understood
as an operator implementing the exterior derivative on the phase-space 
(the ghosts play the role of differentials $c^a \leftrightarrow d\varphi^a$).
The full set of symmetries gives birth to a set of charges whose algebra
is that of $ISp\left(2\right)$. All these charges can be given a
nice geometrical meaning in the framework of Cartan calculus~\cite{Gozzi:1989bf}.

\subsubsection{Super-extension of the Moyal formalism \label{sub:Supersymmetric-extension-of-Moyal-formalism}}

Inspired by the classical case, we study the super-extension of the
Moyal formalism, which has been put forward in~\cite{Gozzi:1993nk} in order
to investigate how to possibly introduce geometric operators, like
the exterior derivative, in QM. Quite generally, given the Hamiltonian
$\widetilde{{\cal H}}_{B}^{\hbar}={\cal F}\left(\lambda_{a}\omega^{ab}\partial_{b}\right)H$,
the authors propose the following super-extension
\begin{eqnarray*}
\widetilde{{\cal H}}^{\hbar} & \equiv & \widetilde{{\cal H}}_{B}^{\hbar}+\widetilde{{\cal H}}_{F}^{\hbar}\\
 & = & {\cal F}\left(\lambda_{a}\omega^{ab}\partial_{b}\right)H+i\bar{c}_{a}\omega^{ac}\partial_{c}\partial_{b}\frac{{\cal F}\left(\lambda_{a}\omega^{ab}\partial_{b}\right)}{\left(\lambda_{a}\omega^{ab}\partial_{b}\right)}Hc^{b}\,,
\end{eqnarray*}
where
\begin{eqnarray*}
{\cal F}\left(\lambda_{a}\omega^{ab}\partial_{b}\right) & \equiv & \frac{2}{\hbar}\left[\sum_{n=0}^{\infty}\frac{1}{\left(2n+1\right)!}\left(\frac{\hbar}{2}\lambda_{a}\omega^{ab}\partial_{b}\right)^{2n+1}\right]\\
 & = & \frac{2}{\hbar}\sinh\left(\frac{\hbar}{2}\lambda_{a}\omega^{ab}\partial_{b}\right)\,,
\end{eqnarray*}
and $\bar{c}_a$ and $c^b$ are Grassmann odd fields, which we call ghosts.

The classical case is recovered via the identification ${\cal F}\left(\lambda_{a}\omega^{ab}\partial_{b}\right)\rightarrow\lambda_{a}\omega^{ab}\partial_{b}$,
consistently with the limit $\hbar\rightarrow0$. As shown in~\cite{Gozzi:1993nk},
the super-extended Hamiltonian $\widetilde{{\cal H}}^{\hbar}$ is determined by the following two requirements:
\begin{enumerate}
\item the complete Hamiltonian $\widetilde{{\cal H}}^{\hbar}$ is assumed
to be invariant under the BRS transformation (\ref{eq:BRS-symmetry-transf});
\item the ghost part of the Hamitonian, $\widetilde{{\cal H}}_{F}^{\hbar}$, is at most
bilinear in the ghosts.
\end{enumerate}
The action associated to the Hamiltonian $\widetilde{{\cal H}}^{\hbar}$
is
\begin{eqnarray}
\int dt\,\widetilde{{\cal L}}^{\hbar} & \equiv & \int dt\,\left[\lambda_{a}\dot{\varphi}^{a}-\widetilde{{\cal H}}_{B}^{\hbar}+i\bar{c}_{a}\dot{c}^{a}-\widetilde{{\cal H}}_{F}^{\hbar}\right]\,.\label{eq:action-for-super-Moyal}
\end{eqnarray}
This action characterizes the path integral weight associated to the
super-extended Moyal formalism, whose partition function is given
by
\begin{eqnarray}
Z & = & \int{\cal D}\varphi^{a}{\cal D}\lambda_{a}{\cal D}c^{a}{\cal D}\bar{c}_{a}\,e^{i\int dt\,\widetilde{{\cal L}}^{\hbar}}\,.\label{eq:Z-via-path-integral-super-Moyal}
\end{eqnarray}
One can readily check that the new action (\ref{eq:action-for-super-Moyal}) is BRS invariant. Moreover,
it is interesting to note that not only the BRS symmetry survives
the ``quantization process'', namely to go from $\widetilde{{\cal H}}^{\hbar=0}$
to $\widetilde{{\cal H}}^{\hbar}$. Rather, all the $ISp\left(2\right)$
symmetry group is still present~\cite{Gozzi:1993nk}. However, the action is no longer
invariant under the $N=2$ supersymmetry present in the classical
case.\footnote{As suggested in~\cite{Gozzi:1993nk} and~\cite{Crossley:2015evo} it would be interesting
to investigate whether a deformed version of the SUSY survives instead.} 

To the super-extended path integral (\ref{eq:Z-via-path-integral-super-Moyal})
one can associate an operatorial formalism, analogous to the one presented
in section \ref{sub:Path-integral-for-WignerMoyal-formalism} (see~\cite{Gozzi:1993nk} for more details). 
In particular, we may consider the representation
\begin{eqnarray*}
\hat{c}^{a}=c^{a} &  & \hat{\bar{c}}_{a}=\frac{\partial}{\partial c^{a}}\,.
\end{eqnarray*}
The path integral can then be thought of as representing the kernel of
propagation associated to the following Schr{\"o}rdinger-like equation 
\begin{eqnarray}
i\partial_{t}\rho\left(\varphi^{a},c^{a};t\right)-\widetilde{{\cal H}}^{\hbar}\rho\left(\varphi^{a},c^{a};t\right) & = & 0\,,\label{eq:super-Schroedinger-like-equation-or-super-Liouville}
\end{eqnarray}
where now the variables in $\widetilde{{\cal H}}^{\hbar}$ are interpreted
as operators. We have already detailed the operatorial form of $\widetilde{{\cal H}}_{B}^{\hbar}$
in section \ref{sub:Path-integral-for-WignerMoyal-formalism}, its
ghost extension acts as follows:
\begin{eqnarray*}
\widetilde{{\cal H}}_{F}^{\hbar}\rho\left(\varphi^{a},c^{a};t\right) & = & \left[i\frac{\partial}{\partial c^{a}}\omega^{ac}\partial_{c}\partial_{b}\frac{{\cal F}\left(\widehat{\lambda}_{a}\omega^{ab}\partial_{b}\right)}{\left(\widehat{\lambda}_{a}\omega^{ab}\partial_{b}\right)}Hc^{b}\right]\rho\left(\varphi^{a},c^{a};t\right)\,.
\end{eqnarray*}

Next, let us show that the ghosts that we just added are harmless
as their contribution corresponds to a unit determinant. Indeed,
the full ghost Lagrangean is given by
\begin{eqnarray*}
\widetilde{{\cal L}}_{F} & = & i\bar{c}_{a}\dot{c}^{b}-i\bar{c}_{a}\omega^{ac}\partial_{c}\partial_{b}\frac{{\cal F}\left(\lambda_{a}\omega^{ab}\partial_{b}\right)}{\left(\lambda_{a}\omega^{ab}\partial_{b}\right)}Hc^{b}\,.
\end{eqnarray*}
Integrating over the ghosts, the Lagrangean $\widetilde{{\cal L}}_{F}$ gives rise to a determinant
in the path integral. This determinant reads 
\begin{eqnarray*}
\int{\cal D}c^{a}{\cal D}\bar{c}_{a}\,e^{i\int\widetilde{{\cal L}}_{F}} & = & \det\left[\left(\partial_{t}-\omega^{ac}\partial_{c}\partial_{b}\frac{{\cal F}\left(\lambda_{a}\omega^{ab}\partial_{b}\right)}{\left(\lambda_{a}\omega^{ab}\partial_{b}\right)}H\right)\delta\left(t-t^{\prime}\right)\right]\\
 & = & \exp\mbox{Tr}\log\left[\partial_{t}\left(1+\partial_{t}^{-1}\omega^{ac}\partial_{c}\partial_{b}\frac{{\cal F}\left(\lambda_{a}\omega^{ab}\partial_{b}\right)}{\left(\lambda_{a}\omega^{ab}\partial_{b}\right)}H\right)\delta\left(t-t^{\prime}\right)\right]\\
 & = & \exp\mbox{tr}\left[\log\left[\left(1+\partial_{t}^{-1}\omega^{ac}\partial_{c}\partial_{a}\frac{{\cal F}\left(\lambda_{a}\omega^{ab}\partial_{b}\right)}{\left(\lambda_{a}\omega^{ab}\partial_{b}\right)}H\right)\delta\left(t-t^{\prime}\right)\right]\right]\\
 & = & \exp\mbox{tr}\left[\log\left[\left(1\right)\delta\left(t-t^{\prime}\right)\right]\right]=1\,,
\end{eqnarray*}
where in the third line we dropped a field independent term. 
(Note that we are considering retarded boundary condition for the Green's function.)

Therefore, if we consider the insertion of operators that do not depend
on the ghosts, like any operator present in the standard Moyal formalism,
we would obtain the same correlation function as in the standard case without ghosts.
This is due to the fact that the integration over the ghosts produces no terms that
depend on the dynamical variables.

Thus, the super-extension of the Moyal formalism is in some sense a redundant
description. However, from the CPI case, we know that, despite
being not strictly needed, the ghost sector considerably enlightens many aspects
of the theory. Indeed, the super-extended version of the Moyal formalism
has been used in~\cite{Gozzi:1993nk} to investigate a possible extension of
Cartan calculus to QM. 

Finally let us recall that, as noted in~\cite{Gozzi:1993nk}, the full Lagrangean
can be written as a BRS-variation, in particular
\begin{eqnarray*}
\widetilde{{\cal L}}^{\hbar} & = & \delta_{{\rm BRS}}\left(\bar{c}_{a}\left(\dot{\varphi}^{a}-\omega^{ab}\partial_{b}
 \frac{{\cal F}\left(\lambda_{a}\omega^{ab}\partial_{b}\right)}{\left(\lambda_{a}\omega^{ab}\partial_{b}\right)}H
\right)\right)\,.
\end{eqnarray*}

Concluding, let us also recall the following rewriting for the Hamiltonian
$\widetilde{{\cal H}}^{\hbar}$ found in~\cite{Gozzi:1993nk}. We define
\begin{eqnarray}
H_{\hbar} & \equiv & \frac{{\cal F}\left(\lambda_{a}\omega^{ab}\partial_{b}\right)}{\left(\lambda_{a}\omega^{ab}\partial_{b}\right)}H\label{eq:def-H_hbar-deformed}\\
 & = & \frac{1}{2}\int_{-1}^{1}ds\,\exp\left[-s\left(\frac{\hbar}{2}\lambda_{a}\omega^{ab}\partial_{b}\right)\right]H\nonumber \\
 & = & \frac{1}{2}\int_{-1}^{1}ds\,H\left(\varphi^{a}-s\frac{\hbar}{2}\lambda_{b}\omega^{ba}\right)\,.\nonumber 
\end{eqnarray}
Using this and introducing $h_{\hbar}^{a}\equiv\omega^{ab}\partial_{b}H_{\hbar}$
we have 
\begin{eqnarray}
\widetilde{{\cal H}}^{\hbar} & = & \lambda_{a}h_{\hbar}^{a}+i\bar{c}_{a}\partial_{b}h_{\hbar}^{a}Hc^{b}\,.\label{eq:Htilde-hbar-via-the-lifted-H}
\end{eqnarray}
At this point we note that the super-extended Hamiltonian of the Moyal formalism, $\widetilde{{\cal H}}^{\hbar}$,
can be obtained from its classical limit by replacing $H\left(\varphi \right)$ with $H_{\hbar} \left(\varphi , \lambda \right)$.

\subsection{Comparison with the Schwinger-Keldysh formalism \label{sub:Comparison-with-the-SK-formalism}}

We wish to brielfly discuss the super-extension of the SK formalism
and its link with the Moyal super-extension. The SK extension has
been studied in detail recently in
\cite{Crossley:2015evo,Glorioso:2016gsa,Gao:2017bqf,Glorioso:2017fpd,Haehl:2015foa,Haehl:2015uoc,Haehl:2016pec,Haehl:2016uah,Haehl:2017zac,Jensen:2017kzi}. 

In~\cite{Crossley:2015evo} the authors construct an effective field theory (EFT) for hydrodynamics by expressing
the low energy path integral in terms of Stuckelberg fields which
allow to keep manifest the symmetries of the original action when equipped
with sources (background gauge fields and the metric). In order to
mantain the unitarity property (\ref{eq:unitarity-condition-SK})
in the low energy EFT, the authors note that introducing certain Grassmann
odd fields (ghosts), and imposing a BRS symmetry analogous to the one
we considered in section \ref{sub:Classical-path-integral} allow
to keep the property (\ref{eq:unitarity-condition-SK}) manifest.
Moreover, it is explicitly checked that this ghost sector is required
in the classical limit. 

On the other hand, in~\cite{Haehl:2016pec} the authors introduce the ghost
sector by considering a generic field redefinition and carrying out
a gauge-fixing like procedure. This procedure does not change the
content of the theory but allows once again to make manifest the unitarity
condition (\ref{eq:unitarity-condition-SK}). The relation between
these two latter frameworks is analyzed in~\cite{Haehl:2017zac}.

In this section we observe that the super-extension
of the path integral for the Moyal approach to QM carries many similarities
with the various super-extensions proposed for the SK formalism. In particular,
in section \ref{sub:Supersymmetric-extension-of-Moyal-formalism}
we have seen that the action (\ref{eq:action-for-super-Moyal}) has
been taylored precisely to inherit the BRS symmetry present in the
classical case. 
A similar reasoning underlies the extension proposed in~\cite{Crossley:2015evo}.
However, in~\cite{Gozzi:1993nk} the ghosts were added with
the purpose of putting forward a path integral implementation of
geometric operations, like the exterior derivative, but essentially
the ghost action has been selected without any strong need behind.
From the SK formalism point of view, we see that there is a further
reason to add the ghost fields, namely to preserve the condition (\ref{eq:unitarity-condition-SK})
in a manifest way~\cite{Crossley:2015evo,Haehl:2016pec}. 

The theory obtained by setting equal the sources on the forward and
backward branches, so that condition (\ref{eq:unitarity-condition-SK})
is realized, is sometimes referred to as the topological sector of
the SK formalism. We wish to show here that also in the Moyal approach
certain topological properties are present. This has been fully investigated
in the classical limit $\hbar\rightarrow0$ in~\cite{Gozzi:1989vv}. We will
prove that similar observations apply also in the quantum case. Before
discussing the quantum case let us recall the classical result. In~\cite{Gozzi:1989vv}
it is shown that the following functional integral describes a topological field theory:
\begin{eqnarray}
Z_{\rm{pbc}} & \equiv & \int_{\rm{pbc}}{\cal D}\varphi^{a}{\cal D}\lambda_{a}{\cal D}c^{a}{\cal D}\bar{c}_{a}\,e^{i\int_{0}^{T}dt\,{\cal L}^{\hbar=0}}\nonumber \\
 & = & \int d\varphi_{0}^{a}dc_{0}^{a}\,K\left(\varphi_{0}^{a},c_{0}^{a},T;\varphi_{0}^{a},c_{0}^{a},0\right)\,,\label{eq:CPI-expression-for-Zpbc}
\end{eqnarray}
where ${\cal L}^{\hbar=0}=\widetilde{{\cal L}}$ is the Lagrangean given in (\ref{eq:CPI-lagrangean}),
and periodic boundary conditions ($\rm{pbc}$) have been imposed on both the phase-space
variables $\varphi^{a}$ and the ghosts $c^{a}$. The fact of having
periodic boundary conditions also for the Grassmann odd fields $c^{a}$
is due to the fact that only in this case the boundary conditions
are also BRS invariant, as is the integrand see~\cite{Gozzi:1989vv}. The next crucial fact proven in~\cite{Gozzi:1989vv}
is that $Z_{\rm{pbc}}$ is actually independent of the dynamics,
i.e.~it does not depend on $H$, and
\begin{eqnarray}
Z_{\rm{pbc}} & = & \mbox{Tr}\left[\left(-1\right)^{F}e^{-i\widetilde{{\cal H}}^{\hbar=0}T} \right]\,=\,\mbox{Tr}\left[\left(-1\right)^{F}\right]\,.\label{eq:Zpbc-and-Witten-index}
\end{eqnarray}
In (\ref{eq:Zpbc-and-Witten-index}) the trace is performed over functions $\rho \left(\varphi^a,c^a \right)$
or, equivalently, a set of forms $\rho_{a_1\cdots a_p} \left(\varphi^a \right)$, with $p\in [0,2n]$.
$F$ indicates the ``fermion'' number operator which counts the degree of the respective differential form,
i.e.~$(-1)^F=+1$ for even $p$ and $(-1)^F=-1$ otherwise.
Thus, $Z_{\rm{pbc}}$ evaluates the Witten index for the CPI. 

In order to generalize these results to the quantum case $\hbar\neq0$
we consider
\begin{eqnarray}
Z_{\rm{pbc}} & \equiv & \int_{\rm{pbc}}{\cal D}\varphi^{a}{\cal D}\lambda_{a}{\cal D}c^{a}{\cal D}\bar{c}_{a}\,e^{i\int_{0}^{T}dt\,{\cal L}^{\hbar}}\label{eq:Zpbc-via-functional-integral-super_Moyal}
\end{eqnarray}
and ask if the value of $Z_{\rm{pbc}}$ in the super-extended Moyal approach depends on $\hbar$:
\begin{eqnarray*}
\frac{d}{d\hbar}Z_{\rm{pbc}} & = & \frac{d}{d\hbar}\int_{\rm{pbc}}{\cal D}\varphi^{a}{\cal D}\lambda_{a}{\cal D}c^{a}{\cal D}\bar{c}_{a}\,e^{i\int_{0}^{T}dt\,{\cal L}^{\hbar}}\,.
\end{eqnarray*}
It is convenient to re-express the path integral via the following
rescaled fields: $\lambda_{a}^{\prime}\equiv\hbar\lambda_{a}$ and
$\bar{c}_{a}^{\prime}\equiv\hbar\bar{c}_{a}$ so that, dropping the
prime, the action reads
\begin{eqnarray*}
\int dt\,\widetilde{{\cal L}}^{\hbar} & = & \frac{1}{\hbar}\int dt\,\left[\lambda_{a}\dot{\varphi}^{a}-\widetilde{{\cal H}}_{B}^{\hbar=1}+i\bar{c}_{a}\dot{c}^{a}-\widetilde{{\cal H}}_{F}^{\hbar=1}\right]\,,
\end{eqnarray*}
where the $\hbar$ is now present only as an overall factor. Let us
point out that the path integral measure is invariant under this rescaling
due to the properties of the Jacobian of Grassmann even and odd variables.
In particular $d\lambda_{a}=d\lambda_{a}^{\prime}/\hbar$ and $d\bar{c}_{a}=d\bar{c}_{a}^{\prime}\hbar$,
owning to the fact that the Jacobian for Grassmann variables is just
the inverse of the standard one. 

As a consequence, when differentiating $Z_{\rm{pbc}}$ with respect to $\hbar$
we find
\begin{eqnarray*}
\frac{d}{d\hbar}Z_{\rm{pbc}} & = & \int_{\rm{pbc}}{\cal D}\varphi^{a}{\cal D}\lambda_{a}{\cal D}c^{a}{\cal D}\bar{c}_{a}\,e^{\frac{i}{\hbar}\int_{0}^{T}dt\,{\cal L}^{\hbar=1}}\left(-\frac{i}{\hbar^{2}}\int_{0}^{T}dt^{\prime}\,{\cal L}^{\hbar=1}\right)\,.
\end{eqnarray*}
Since the Lagrangian can be re-expressed as BRS variation we have
that
\begin{eqnarray*}
\frac{d}{d\hbar}Z_{\rm{pbc}} & = & -\frac{i}{\hbar^{2}}\int_{0}^{T}dt^{\prime}\,\int_{\rm{pbc}}{\cal D}\varphi^{a}{\cal D}\lambda_{a}{\cal D}c^{a}{\cal D}\bar{c}_{a}\,e^{\frac{i}{\hbar}\int_{0}^{T}dt\,{\cal L}^{\hbar=1}}\left[\delta_{{\rm BRS}}\left(\bar{c}_{a}\left(\dot{\varphi}^{a}-\omega^{ab}\partial_{b}H_{\hbar}\right)\right)\right]\,,
\end{eqnarray*}
which, upon integrating by parts in field space, yields a vanishing
result due to the nilpotent character of the BRS transformation, $\left(\delta_{\rm{BRS}}\right)^2=0$. 

One can also give a further argument that suggests that $Z_{\rm{pbc}}$ is actually
independent of $\hbar$. The topological nature of $Z_{\rm{pbc}}$ for
the CPI is tied to its invariance with respect to arbitrary deformations of
the Hamiltonian, $H(\varphi^a)\rightarrow H(\varphi^a)+\delta H(\varphi^a)$. However, it is straightforward to check that the super-extended
Moyal Lagrangean ${\cal L}^\hbar$ can be obtained from the CPI one, i.e.~$\widetilde{\cal L}$,
by replacing $H$ with $H_{\hbar}$,
introduced in equation (\ref{eq:def-H_hbar-deformed}). Since, the
classical $Z_{\rm{pbc}}$ is invariant under deformation of $H$ it is
also invariant under the deformation $H\rightarrow H_{\hbar}$.
Note, however, that $ H_{\hbar}$ depends also on $\lambda_a$, and not only on $\varphi^a$, and so, strictly speaking,
the deformation $H\rightarrow H_{\hbar}$ does not belong to class $H(\varphi^a)\rightarrow H(\varphi^a)+\delta H(\varphi^a)$
that we are considering. Therefore, this latter observation is just an argument, 
the proof of our previous statements is actually based on the BRS invariance.

To make the discussion self-contained, let us mention how to obtain $Z_{\rm{pbc}}$. 
All the details can be found
in ref.~\cite{Gozzi:1989vv}. Since $Z_{\rm{pbc}}$ is independent from $\hbar$, we
set $\hbar=0$ and work with the CPI Lagrangean. Due to equation (\ref{eq:Zpbc-and-Witten-index}),
$Z_{\rm{pbc}}$ is actually independent of the time $T$ and we can compute it via
the associated path integral (\ref{eq:CPI-expression-for-Zpbc})
in the limit $T\rightarrow0$. In the CPI Lagrangean the fields $\lambda_{a}$
and $\bar{c}_{a}$ appear linearly and functionally integrating them
out simply produces a delta function. Hence, using the linearized
solution of the equation of motion in the limit $T\rightarrow0$ we
have~\cite{Gozzi:1989vv}
\begin{eqnarray*}
Z_{\rm{pbc}} & = & \int d\varphi_{0}^{a}dc_{0}^{a}\,\delta\left(\varphi_{0}^{a}-\left(\varphi_{0}^{a}+\omega^{ab}\partial_{b}HT\right)\right)\delta\left(c_{0}^{a}-\left(c_{0}^{a}+\omega^{ac}\partial_{c}\partial_{b}HTc_{0}^{b}\right)\right)\\
 & = & \sum_p \int d\varphi_{0}^{a}dc_{0}^{a}\,\frac{\det\left(\omega^{ac}\partial_{c}\partial_{b}H\left(\varphi_{\left(p\right)}\right)\right)}{\left|\det\left(\omega^{ac}\partial_{c}\partial_{b}H\left(\varphi_{\left(p\right)}\right)\right)\right|}\delta\left(\varphi_{0}-\varphi_{\left(p\right)}\right)\delta\left(c_{0}^{b}\right)\\
 & = &  \sum_p \left(-1\right)^{i_p} \, = \, \chi\left({\cal M}_{2n}\right)\,.
\end{eqnarray*}
In the above lines $\varphi_{\left(p\right)}$ are the critical points of $H$, i.e.~those points in which the Hamiltonian vector field vanishes,
while $i_p$ is the number of negative eigenvalues of the Hessian of $H$ at the critical point.
It follows that $Z_{\rm{pbc}}$ is actually the Morse theory representation of the Euler number of the manifold
on which the ``Morse function'' $H$ is defined, i.e.~the phase-space
manifold associated to the mechanical system~\cite{Gozzi:1989vv}. Along the lines of ref.~\cite{Gozzi:1989vv},
it is also possible to study other topological observables.

Therefore, we have shown that there exists a topological sector also in the super-extended
Moyal formalism. Despite being seemingly close, the topological sector
that we have just discussed and that usually considered in the SK formalism
are not the same. To see this we recall that the topological sector
in the SK formalism is related to the unitarity condition (\ref{eq:unitarity-condition-SK}),
which can be rewritten as
\begin{eqnarray}
\mbox{Tr}\left[U\left(t,0;J\right)\rho_{0}U^{\dagger}\left(t,0;J\right)\right] & = & \mbox{Tr}\left[\rho_{0}\right]\,,\label{eq:unitarity-condition-topological-source-dependence}
\end{eqnarray}
whose RHS is explicitly independent from the Hamiltonian evolving
the system. Therefore, also the path integral representation of (\ref{eq:unitarity-condition-topological-source-dependence})
is independent from the dynamics of the system. However, as we shall
check in a moment, the path integral expression associated to (\ref{eq:unitarity-condition-topological-source-dependence})
is different from the one concerning $Z_{\rm{pbc}}$. We have
\begin{eqnarray*}
\mbox{Tr}\left[U\left(t,0;J\right)\rho_{0}U^{\dagger}\left(t,0;J\right)\right] & = & \mbox{Tr}\left[\rho_{t}\right]\\
 & = & \int d\varphi_{f}^{a}dc_{f}^{a}\,\rho\left(\varphi_{f}^{a},c_{f}^{a};t\right)\\
 & = & \int d\varphi_{f}^{a}dc_{f}^{a}d\varphi_{0}^{a}dc_{0}^{a}\,K\left(\varphi_{f}^{a},c_{f}^{a},t;\varphi_{0}^{a},c_{0}^{a},0\right)\rho\left(\varphi_{0}^{a},c_{0}^{a};0\right)\,,
\end{eqnarray*}
where the kernel of propagation $K\left(\varphi_{f}^{a},c_{f}^{a},t;\varphi_{0}^{a},c_{0}^{a},0\right)$
is given in terms of the super-extended Moyal path integral. A particularly
simple choice for the initial density $\rho\left(\varphi_{0}^{a},c_{0}^{a};0\right)$
is
\begin{eqnarray*}
\rho\left(\varphi_{0}^{a},c_{0}^{a};0\right) & = & \delta\left(\varphi_{0}^{a}-\varphi^{a}\right)\delta\left(c_{0}^{a}-c^{a}\right)\,.
\end{eqnarray*}
With this choice we have the following path integral representation
\begin{eqnarray*}
\mbox{Tr}\left[U\left(t,0;J\right)\rho_{0}U^{\dagger}\left(t,0;J\right)\right] & = & \int d\varphi_{f}^{a}dc_{f}^{a}\,K\left(\varphi_{f}^{a},c_{f}^{a},t;\varphi^{a},c^{a},0\right)\\
 & = & \int d\varphi_{f}^{a}dc_{f}^{a}\,\int_{\varphi ,c}^{\varphi_{f},c_{f}}{\cal D}\varphi^{a}{\cal D}\lambda_{a}{\cal D}c^{a}{\cal D}\bar{c}_{a}\,e^{i\int_{0}^{t}dt\,{\cal L}^{\hbar}}\,.
\end{eqnarray*}
This latter expression is clearly different from the one concerning
$Z_{\rm{pbc}}$ due to the different boundary conditions. To see this more
explicitly we can mimic the reasoning that we did for $Z_{\rm{pbc}}$.
Since we know that actually the topological SK sector is independent
of $\hbar$ due to equation (\ref{eq:unitarity-condition-topological-source-dependence}),
at least whenever $\hbar$ does not enter in the initial density matrix
as in our case, we can set $\hbar=0$ and work once again with the
CPI. Integrating over $\lambda_{a}$ and $\bar{c}_{a}$ produces again
a delta function, and in the limit $t\rightarrow0$
\begin{eqnarray*}
\mbox{Tr}\left[U\left(t,0;J\right)\rho_{0}U^{\dagger}\left(t,0;J\right)\right] & = & \int d\varphi_{f}^{a}dc_{f}^{a}\,\delta\left(\varphi_{f}^{a}-\left(\varphi^{a}+\omega^{ab}\partial_{b}Ht\right)\right) \\
& & \times \delta\left(c_{f}^{a}-\left(c^{a}+\omega^{ac}\partial_{c}\partial_{b}HTc^{b}\right)\right)\\
 & = & 1\,,
\end{eqnarray*}
where we essentialy found the normalization condition $\mbox{Tr}\left[\hat{\rho}_{t}\right]=\mbox{Tr}\left[\hat{\rho}_{0}\right]=1$,
as it should be.

Next, let us discuss a further aspect concerning
the super-extended Moyal approach of section \ref{sub:Supersymmetric-extension-of-Moyal-formalism}
and the super-extended SK formalism. In section \ref{sec:Schwinger-Keldysh-formalism-and-ph-space-PI}
we have seen that the actions of the Moyal path integral and the SK
one coincide after a suitable redefinition of the fields. It follows
that any symmetry present in the SK formulation can be translated
into an associated symmetry transformation in the Moyal formalism.
As we mentioned at the end of section \ref{sec:Schwinger-Keldysh-formalism-and-ph-space-PI},
the KMS condition (\ref{eq:KMS-condition-on-W-for-cpt-invariant-case})
can be seen as a consequence of the time reversal symmetry on the
forward branch and the time reversal plus an imaginary time translation
in the backward branch (provided that the action is time reversal invariant, see e.g.~\cite{Crossley:2015evo,Jensen:2017kzi}). 
This symmetry can be translated also in the
Moyal language, for instance $q\left(t\right)=\frac{1}{2}\left(x_{1}\left(t\right)+x_{2}\left(t\right)\right)$
is mapped to $\frac{1}{2}\left(x_{1}\left(-t\right)+x_{2}\left(-t-i\beta\right)\right)=\frac{1}{4}\left(-\lambda_{q}\left(-t\right)+\lambda_{q}\left(-t-i\beta\right)+2q\left(-t\right)+2q\left(-t-i\beta\right)\right)$.
The crucial fact that allows to translate a SK symmetry in the Moyal framework
is that the Moyal action $S_{{\rm M}}$ and the SK action $S_{{\rm SK}}$
are mapped one onto the other via the mapping discussed in section
\ref{sec:Schwinger-Keldysh-formalism-and-ph-space-PI}. In particular,
via a suitable redefinition of the field, we have seen that the action
$S_{{\rm M}}$ can be ``split'' into two parts, which correspond
to those associated to the backward and forward branches in the SK
formalism. It is natural to ask whether this splitting property is
enjoyed also by the ghost sector.
Seemingly, this is not the case as 
the part of the action containing $\widetilde{{\cal H}}_{F}^{\hbar}$
can not be split along the lines used for $\widetilde{{\cal H}}_{B}^{\hbar}$,
at least straightforwardly. 

The reader may worry that the KMS condition is lost even at the level
of correlation functions. This is not the case for observables which
do not depend on the ghosts. Indeed we have shown that the ghost integration
is actually harmless and produces just a factor of unity in the path integral.
Because of this, the KMS condition is still satisfied at the level
of correlation functions.

With regard to the KMS condition, a very interesting observation applies
to the classical limit $\hbar\rightarrow0$. In~\cite{Gozzi:1989xz} it was shown
that actually the classical limit of the KMS condition can be derived
from the CPI Lagrangean as a Ward identity associated to the supersymmetry
present in the CPI. However, the supersymmetry present in the CPI
is no longer present in the quantum case and it has been speculated
that a quantum version of it might be present~\cite{Gozzi:1993nk,Crossley:2015evo,Gao:2017bqf}.
Recently, a recipe to write down supersymmetric out of equilibrium
field theories has been put forward~\cite{Gao:2017bqf}, it would be interesting
to investigate whether or not this is applicable to the Moyal case.
If this were the case, one would have a new proposal for the super-extension
of the theory and possibly have a new way to define geometric operations
such as exterior derivatives in QM.

Finally, it would be interesting to look in more detail at the relation
between the KMS condition as arising in the path integral formulation
of the Moyal formalism and that developed in deformation quantization,
which introduces the notion of conformal symplectic structure~\cite{confSymplStru}.

\section{BBGKY hierarchy \label{sec:BBGKY-hierarchy}}

So far we have been discussing the path integral associated to the
evolution of the phase-space density distribution $\rho\left(\varphi\right)$.
However, $\rho\left(\varphi\right)$ often carries much more information
than what one is interested in. In kinetic theory the reduced distribution
functions (defined below) play a very important role. The reduced
distribution functions $f_{n}$ for a system with $N$ degrees of freedom are defined by
\begin{eqnarray*}
f_{n}\left(\varphi_{1},\cdots\varphi_{n};t\right) & \equiv & \frac{N!}{\left(N-n\right)!}\int d\varphi_{n+1}\cdots d\varphi_{N}\,\rho\left(\varphi^{a};t\right)\,,
\end{eqnarray*}
where $d\varphi_{i}\equiv dq_{i}dp_{i}$. Eventually, we will denote
$d\Gamma$ the integration element over the full phase-space, while
the integration element over the phase-space of $n$ particles only
will be indicated with $d\Gamma_{n}$.

In classical mechanics the functions $f_{n}$ satisfy an hierarchy
of equations, the BBGKY hierarchy, that describes the temporal evolution
of $f_{n}$ via a partial differential equation involving $f_{n}$
and $f_{n+1}$, giving rise to an infinite hierarchy of equations.
In the following we deduce the BBGKY hierarchy via path integral methods
connected to the Moyal approach to QM. We shall first deal with the
bosonic, i.e.~without ghost, case and then consider the full super-extended framework.

\subsection{Bosonic BBGKY hierarchy \label{sub:Bosonic-BBGKY-hierarchy}}

In order to derive the BBGKY hierarchy in our framework, we need first
to consider a suitable path integral description. 
For formal manipulations it turns out convenient to consider
the path integral implementing the kernel of propagation directly on $\rho$.
The evolution of $\rho$ is determined by equation (\ref{eq:quantum-Liouville-equation})
and its kernel of propagation can be given a path integral representation
with the techniques reviewed in section \ref{sub:Classical-path-integral}.
This program has been implemented successfully in the classical case
in~\cite{JolicoeurGuillou,Gozzi:1990pf,GozziReuterBBGKY}, in this section we extend this program to the quantum
case.

In practice, $\rho$ becomes a field and the phase-space variables
are integrated over in the associated action. 
Thus, we consider the following functional integral:
\begin{eqnarray}
\widetilde{Z} & = & \int{\cal D}\Lambda{\cal D}\rho\,\exp\left[i\int dtd\Gamma\Lambda\left(\partial_{t}-\hat{L}_\hbar\right)\rho\right]\,.\label{eq:second-quantized-Moyal-path-integral}
\end{eqnarray}
Since equation (\ref{eq:quantum-Liouville-equation}), which determines
the evolution of $\rho$, is linear in $\rho$ then its kernel of
propagation is simply quadratic in $\rho$ and its associate ``conjugate''
field $\Lambda$. Furthermore, following~\cite{JolicoeurGuillou,Gozzi:1990pf,GozziReuterBBGKY}, we can generically parametrize the
field $\Lambda$ as follows:
\begin{eqnarray*}
\Lambda\left(\varphi^{a},t\right) & = & \Lambda_{0}\left(t\right)+\left(\sum_{i=1}^{N}\Lambda_{1}\left(q_{i},p_{i},t\right)\right)+\left(\sum_{i<j}\Lambda_{2}\left(q_{i},p_{i},q_{j},p_{j},t\right)\right)+\cdots\,.
\end{eqnarray*}
Note that since the particles are indistinguishable the functions
$\rho$, $f_{n}$ and the fields $\Lambda_{i}$ are totally symmetric.
The action is now characterized by the following structure:
\begin{eqnarray}
\int\Lambda\left(\partial_{t}-\widehat{L}_{\hbar}\right)\rho & = & \int dtd\Gamma \, \Lambda_{0}\left(t\right)\left(\partial_{t}+\widehat{L}_{\hbar}\right)\rho\nonumber \\
 &  & +\int dtd\Gamma\left[\sum_{i=1}^{N}\Lambda_{1}\left(q_{i},p_{i};t\right)\left(\partial_{t}+\widehat{L}_{\hbar}\right)\rho\right]\nonumber \\
 &  & +\int dtd\Gamma\left[\sum_{i<j}^{N}\Lambda_{2}\left(q_{i},p_{i},q_{j},p_{j};t\right)\left(\partial_{t}+\widehat{L}_{\hbar}\right)\rho\right]+\cdots\,.\label{eq:second-quantized-Moyal}
\end{eqnarray}
The operator $\widehat{L}_{\hbar}$, introduced in equation (\ref{eq:quantum-Liouville-equation}),
contains derivatives with respect to the phase-space variables acting
on $\rho$. Now we wish to show that the part of the action identified
by the field $\Lambda_{n}\left(q_{i_{1}},p_{i_{1}},\cdots,q_{i_{n}},p_{i_{n}};t\right)$
reproduces the quantum version of the equation for $\rho_{n}$ in
the BBGKY hierarchy. 

First let us specify the Hamiltonian explicitly:
\begin{eqnarray*}
H & \equiv & H_{n}+H_{N-n}+H_{\rm{m}}\,,
\end{eqnarray*}
where we distinguished the following three contributions
\begin{eqnarray*}
H_{n} & = & \sum_{i=1}^{n}\frac{p_{i}^{2}}{2m}+U\left(q_{i}\right)+\frac{1}{2}\sum_{i\neq j=1}^{n}V\left(q_{i}-q_{j}\right)\\
H_{N-n} & = & \sum_{i=n+1}^{N}\frac{p_{i}^{2}}{2m}+U\left(q_{i}\right)+\frac{1}{2}\sum_{i\neq j=n+1}^{N}V\left(q_{i}-q_{j}\right)\\
H_{{\rm m}} & = & \sum_{i<j}V\left(q_{i}-q_{j}\right)\,.
\end{eqnarray*}
The Hamiltonians $H_{n}$ and $H_{N-n}$ describes respectively the
interaction of two subsets of $n$ and $N-n$ particles with themselves.
The term $H_{{\rm m}}$, where the subscript ${\rm m}$ stays for
mixed, contains the interaction between the $N-n$ particles, that
we wish to integrate over in $\rho$, and the remaining $n$ particles
(the sum in $H_{{\rm m}}$ is ordered for practical purposes).

Let us consider that the operator $\widehat{L}_{B}^{\hbar}\equiv\widehat{L}_{\hbar}$
is linear in $H$ (the subscript $B$ is introduced for later purposes).
This implies that we can easily separate the three contributions and
distinguish the three operators $\widehat{L}_{B}^{\hbar}\left(H_{n}\right)$,
$\widehat{L}_{B}^{\hbar}\left(H_{N-m}\right)$ and $\widehat{L}_{B}^{\hbar}\left(H_{{\rm m}}\right)$.
We shall focus on the term containing the field $\Lambda_{n}$ in
the action (\ref{eq:second-quantized-Moyal}) and show that it yields
a suitable quantum version of the BBKGY hierarchy. In practice, we
focus on the term
\begin{eqnarray}
\int{\cal L}_{n} & \equiv & \int dtd\Gamma \,
\frac{1}{n!}\Biggr[\frac{N!}{\left(N-n\right)!}\Lambda_{n}\left(q_{1},\cdots,q_{n},p_{1},\cdots,p_{n};t\right)\Biggr(\partial_{t}+\widehat{L}_{B}^{\hbar}\left(H_{n}\right)\nonumber \\
 &  & +\widehat{L}_{B}^{\hbar}\left(H_{N-n}\right)+\widehat{L}_{B}^{\hbar}\left(H_{{\rm m}}\right)\Biggr)\Biggr]\rho\,,\label{eq:Lagrangian-for-n-reduced-distribution-bosonic-case}
\end{eqnarray}
where we took advantage of the totally symmetric character of the
terms in the action. We note that a functional integration
over the field $\Lambda_{n}$ produces a delta function in the path
integral. In the following we will show that the expression contained
in this delta function yields precisely the equation for $f_{n}$
in the BBGKY hierarchy.

Let us consider each of the terms appearing in ${\cal L}_{n}$ of (\ref{eq:Lagrangian-for-n-reduced-distribution-bosonic-case}). We have
\begin{eqnarray*}
\int dt d\Gamma \,
\frac{1}{n!}\Biggr[\frac{N!}{\left(N-n\right)!}\Lambda_{n}\partial_{t}\rho\Biggr] & = & \int dtd\Gamma_{n}\frac{1}{n!}\Biggr[\frac{N!}{\left(N-n\right)!}\Lambda_{n}\partial_{t}\left(\int d\Gamma_{N-n}\rho\right)\Biggr]\\
 & = & \int dtd\Gamma_{n}\frac{1}{n!}\Biggr[\Lambda_{n}\partial_{t}f_{n}\Biggr]\,,
\end{eqnarray*}
where we exploited the fact that the field $\Lambda_{n}$ does not
depend on the phase-space variables of the $N-n$ particle set. Thus,
we note the natural appearence of the reduced distribution $f_{n}$.

The second term in (\ref{eq:Lagrangian-for-n-reduced-distribution-bosonic-case}) can be dealt with in full analogy. Indeed, the Hamiltonian
$H_{n}$ contains by definition only phase-space variables associated
to the set having $n$ particles. We then have
\begin{eqnarray*}
\int dtd\Gamma \,
\frac{1}{n!}\Biggr[\frac{N!}{\left(N-n\right)!}\Lambda_{n}\widehat{L}_{B}^{\hbar}\left(H_{n}\right)\Biggr]\rho & = & \int dtd\Gamma_{n}\frac{1}{n!}\Biggr[\Lambda_{n}\widehat{L}_{B}^{\hbar}\left(H_{n}\right)f_{n}\Biggr]\,.
\end{eqnarray*}

Let us turn now to the third term in equation (\ref{eq:Lagrangian-for-n-reduced-distribution-bosonic-case}).
The operator $\widehat{L}_{B}^{\hbar}\left(H_{N-n}\right)$ reads
\begin{eqnarray*}
\widehat{L}_{B}^{\hbar}\left(H_{N-n}\right) & = & -\frac{2}{\hbar}\sum_{m=0}^{\infty}\frac{\left(-1\right)^{m}}{\left(2m+1\right)!}\left(\frac{\hbar}{2}\right)^{2m+1}\omega^{a_{1}b_{1}}\cdots\omega^{a_{2m+1}b_{2m+1}}\partial_{a_{1}}\cdots\partial_{a_{2m+1}}H_{N-n} \\ 
&\, & \times \partial_{b_{1}}\cdots\partial_{b_{2m+1}}\,.
\end{eqnarray*}
In ${\cal L}_{n}$, the derivatives contained in $\widehat{L}_{B}^{\hbar}\left(H_{N-n}\right)$
act on $\rho$. Given that $H_{N-n}$ depends only on the $N-n$ particles,
we note that the ``effective'' derivatives acting on $H_{N-n}$
are those with respect to these $N-n$ particles. This determines
which derivatives act on $\rho$. For instance, let $a_{i}=p_{i}$
then in $\widehat{L}_{B}^{\hbar}\left(H_{N-n}\right)$ we have a term
proportional to $-\partial_{p_{i}}H_{N-n}\partial_{q_{i}}$. Now we
note that we can integrate by parts in ${\cal L}_{n}$ the derivatives
acting on $\rho$ so that the term $-\partial_{p_{i}}H_{N-n}\partial_{q_{i}}$
becomes $\partial_{q_{i}}\partial_{p_{i}}H_{N-n}=0$. Therefore, we
have shown that the term coming from $\widehat{L}_{B}^{\hbar}\left(H_{N-n}\right)$
actually vanishes.

Finally, we turn to the term in (\ref{eq:Lagrangian-for-n-reduced-distribution-bosonic-case})
due to $\widehat{L}_{B}^{\hbar}\left(H_{{\rm m}}\right)$, which reads
\begin{eqnarray*}
\widehat{L}_{B}^{\hbar}\left(H_{{\rm m}}\right) & = & -\frac{2}{\hbar}\sum_{m=0}^{\infty}\frac{\left(-1\right)^{m}}{\left(2m+1\right)!}\left(\frac{\hbar}{2}\right)^{2m+1}\omega^{a_{1}b_{1}}\cdots\omega^{a_{2m+1}b_{2m+1}}\partial_{a_{1}}\cdots\partial_{a_{2m+1}}H_{{\rm m}} \\
&\, & \times \partial_{b_{1}}\cdots\partial_{b_{2m+1}}\,.
\end{eqnarray*}
Let us focus on a single term of the sum. If one of the derivatives
acting on $H_{{\rm m}}$ is performed with respect to one of the phase-space
variables of the $N-n$ particles set then, by the same reasoning
used for $H_{N-n}$, the contribution vanishes. Therefore, the only
non-vanishing contributions are those in which all the derivative
acting on $H_{{\rm m}}$ are performed with respect to the $n$ phase-space
variables of the $n$ particles set. Such contributions are proportional
to
\begin{eqnarray*}
\partial_{a_{1}}\cdots\partial_{a_{2m+1}}H_{{\rm m}} & = & \partial_{a_{1}}\cdots\partial_{a_{2m+1}}\left(\sum_{i=1}^{n}\sum_{j=n+1}^{N}V\left(q_{i}-q_{j}\right)\right)\,,
\end{eqnarray*}
where the derivatives acts only on the $q_{i}$ dependence. Given
the totally symmetric character of $\rho$ and $\Lambda_{n}$ we can
relabel each term in $\sum_{j=n+1}^{N}V\left(q_{i}-q_{j}\right)$
in the action via $\sum_{j=n+1}^{N}V\left(q_{i}-q_{j}\right)=\left(N-n\right)V\left(q_{i}-q_{n+1}\right)$.
Once rewritten in this way, we note that the dependence on the phase-space
variables $\left\{ q_{i},p_{i}\right\} _{i=n+2,\cdots,N}$ in the
terms of the action due to $\widehat{L}_{B}^{\hbar}\left(H_{{\rm m}}\right)$
is contained only in $\rho$. Thus, we can already forecast that by
integrating over these latter variables we will determine
a contributions proportional to $f_{n+1}$. More precisely: 
\begin{eqnarray*}
\int dtd\Gamma \,
\frac{1}{n!}\Biggr[\frac{N!}{\left(N-n\right)!}\Lambda_{n}\widehat{L}_{B}^{\hbar}\left(H_{{\rm m}}\right)\Biggr]\rho 
& = & 
\int dt d\Gamma_{n+1}\frac{\Lambda_{n}}{n!}\Biggr[-\frac{2}{\hbar}\sum_{m=0}^{\infty}\frac{\left(-1\right)^{m}}{\left(2m+1\right)!}\left(\frac{\hbar}{2}\right)^{2m+1}\\
 &  & \times\omega^{a_{1}b_{1}}\cdots\omega^{a_{2m+1}b_{2m+1}} \\
 &  & \times \partial_{a_{1}}\cdots\partial_{a_{2m+1}}\left(\sum_{i=1}^{n}V\left(q_{i}-q_{n+1}\right)\right) \\
 & &  \times\partial_{b_{1}}\cdots\partial_{b_{2m+1}}f_{n+1}\Biggr]\,.
\end{eqnarray*}

As we anticipated, by performing a functional integration over $\Lambda_{n}$
a (functional) Dirac delta is generated implying
\begin{eqnarray}
\partial_{t}f_{n}+\widehat{L}_{B}^{\hbar}\left(H_{n}\right)f_{n} & = & \sum_{i=1}^{n}\int dq_{n+1}dp_{n+1}\left[-\widehat{L}_{B}^{\hbar}\left(V\left(q_{i}-q_{n+1}\right)\right)f_{n+1}\right]\,.\label{eq:bosonic-BBGKY-via-operator-L}
\end{eqnarray}
The above expression tells us that the evolution of the reduced density
$f_{n}$ depends on the reduced density $f_{n+1}$, thus giving rise
to an infinite hierarchy. Equation (\ref{eq:bosonic-BBGKY-via-operator-L})
can be rewritten in term of Moyal brackets as
\begin{eqnarray}
\partial_{t}f_{n}-\left\{ H_{n},f_{n}\right\} _{mb} & = & \sum_{i=1}^{n}\int dq_{n+1}dp_{n+1}\left\{ V\left(q_{i}-q_{n+1}\right),f_{n+1}\right\} _{mb}\,.\label{eq:BBGKY-eq-via-Moyal-brackets-bosonic}
\end{eqnarray}
By using the isomorphism between the Moyal approach and the operatorial
approach to QM we can further write
\begin{eqnarray*}
i\hbar\partial_{t}\hat{f}_{n}-\left[H_{n},\hat{f}_{n}\right] & = & \sum_{i=1}^{n}\mbox{Tr}_{n+1}\left[\widehat{V}\left(q_{i}-q_{n+1}\right),\hat{f}_{n+1}\right]\,,
\end{eqnarray*}
which is the standard form of the quantum version of the BBGKY hierarchy~\cite{quantum-kinetic-theory}. 

Furthermore, equation (\ref{eq:BBGKY-eq-via-Moyal-brackets-bosonic})
is seen to go into its classical limit simply by replacing the Moyal
brackets with the Poisson brackets or, equivalently, setting $\hbar\rightarrow0$.
Clearly, it is simple to keep the first non-trivial quantum corrections
by truncating at some order in $\hbar$ the expression for the Moyal
brackets.

Finally we point out that, besides deriving the BBGKY hierarchy in a novel way,
our formalism gives an handy framework to be used in kinetic theory.
Indeed, the classical limit of the field theory (\ref{eq:second-quantized-Moyal-path-integral})
has been used in~\cite{JolicoeurGuillou} to derive in an easy way the Balescu-Lenard
collision operator of a high temperature plasma. Further applications have been investigated 
in~\cite{Chavanis1,Chavanis2}.

\subsection{Super BBGKY hierarachy \label{sub:Super-BBGKY-hierarachy}}

In this section we generalize our findings to the super-extended framework.
In this case $\rho$ also depends on the ghost fields and has the
following structure:\footnote{The higher forms of $\rho$ are defined with an unrestricted index summation intended,
hence the factorial in the denominator. Alternatively
an ordering must be chosen, for instance
\[
\frac{1}{\left(2N\right)!}\rho_{d_{1}\cdots d_{2N}}\left(\varphi^{a}\right)c^{d_{1}}\cdots c^{d_{2N}}=\rho_{q_{1}p_{1}\cdots q_{N}p_{N}}\left(\varphi^{a}\right)c^{q_{1}}c^{p_{1}}\cdots c^{q_{N}}c^{p_{N}}\,.
\]
}
\begin{eqnarray*}
\rho\left(\varphi^{a},c^{a}\right) & = & \rho_{0}\left(\varphi^{a}\right)+\rho_{d}\left(\varphi^{a}\right)c^{d}+\cdots+\frac{1}{\left(2N\right)!}\rho_{d_{1}\cdots d_{2N}}\left(\varphi^{a}\right)c^{d_{1}}\cdots c^{d_{2N}}\,.
\end{eqnarray*}
The zero form density $\rho_{0}\left(\varphi^{a}\right)$ corresponds
to the standard density $\rho\left(\varphi\right)$ that we were considering
for the bosonic theory. 
Indeed, it can be checked that $\rho_{0}\left(\varphi^{a}\right)$
is actually determined by (\ref{eq:quantum-Liouville-equation})
given that the terms in (\ref{eq:super-Schroedinger-like-equation-or-super-Liouville}) due to ${\cal H}_{F}^{\hbar}$ vanish automatically
on zero-forms. 
However, also the $2N$-form can be identified with
the ordinary density $\rho\left(\varphi\right)$ once the ghosts are
integrated out. In fact, the operator ${\cal H}_{F}^{\hbar}$ vanishes
automatically also on $2N$-forms so that its coefficient is simply
determined by the bosonic sector, recovering thus the standard density
via $\frac{1}{\left(2N\right)!}\rho_{d_{1}\cdots d_{2N}}\left(\varphi^{a}\right)c^{d_{1}}\cdots c^{d_{2N}}=\rho\left(\varphi\right)c^{q_{1}}\cdots c^{p_{N}}$.

We define the super reduced density distributions as follows:
\begin{eqnarray*}
f_{n}\left(\varphi^{a},c^{a};t\right) & \equiv & \frac{N!}{\left(N-s\right)!}\int d\Gamma_{N-n}d\Gamma_{N-n}^{{\rm g}}\rho\left(\varphi^{a},c^{a};t\right)\,,
\end{eqnarray*}
where $d\Gamma_{N-n}^{{\rm g}}$ denotes the integration element over
the ghost associated to the particles in the $N-n$ set. Equivalently,
we can also introduce the super-coordinate $\chi_{i}\equiv\left(q_{i},p_{i},c^{q_{i}}\equiv\xi_{i},c^{p_{i}}\equiv\pi_{i}\right)$
and write
\begin{eqnarray*}
f_{n}\left(\chi_{1},\cdots ,\chi_{n};t\right) & = & \frac{N!}{\left(N-n\right)!}\int d\chi_{n+1}\cdots d\chi_{N}\,\rho\left(\chi_{1},\cdots,\chi_{N};t\right)\,.
\end{eqnarray*}

The first observation is that also the super-Liouville operator encoded
in $\widetilde{{\cal H}}^{\hbar}$ is linear in the Hamiltonian $H$.
Thus, we can again separate this operator into three terms associated
to $H_{n}$, $H_{N-n}$, and $H_{{\rm m}}$, respectively. In particular,
we consider
\begin{eqnarray*}
L_{F}^{\hbar}\rho \left(\chi ;t\right)
& = & i\widetilde{{\cal H}}_{F}^{\hbar}\rho \left(\chi ;t\right)\\
& = & \left[\omega^{ac}\partial_{c}\partial_{b}\frac{2}{\hbar}\left(\sum_{n=0}^{\infty}\frac{\left(-1\right)^{n}}{\left(2n+1\right)!}\left(\frac{\hbar}{2}\right)^{2n+1}\left(\omega^{ab}\partial_{b}^{\left(H\right)}\partial_{a}^{\left(\rho\right)}\right)^{2n}\right)Hc^{b}\frac{\partial}{\partial c^{a}}\right]\rho \left(\chi ;t\right). \,\,\,
\end{eqnarray*}

Having already discussed the properties of the operator $L_{B}^{\hbar}$
in section \ref{sub:Bosonic-BBGKY-hierarchy}, let us move to 
those of $L_{F}^{\hbar}$. It is straightforward to check that $L_{F}^{\hbar}\left(H_{n}\right)$
and $L_{F}^{\hbar}\left(H_{N-n}\right)$ depend only on the $n$
and $N-n$ $\chi$-variables respectively, and so the derivatives
appearing in these operators are effectively acting only on such variables.
Because of this, in the very same way as we discussed in section \ref{sub:Bosonic-BBGKY-hierarchy},
the operator $L_{F}^{\hbar}\left(H_{H-n}\right)$ disappears from the
part of the action proportional to $\Lambda_{n}$. Looking at the
part of the action where $L_{F}^{\hbar}\left(H_{n}\right)$ appear,
we note that the variables $\chi_{n+1},\cdots,\chi_{N}$ only appear
in the density $\rho$. Thus, we can integrate over them and conclude
that $L_{F}^{\hbar}\left(H_{n}\right)$ acts on the reduced density
$f_{n}$. 

Next, let us consider $L_{F}^{\hbar}\left(H_{{\rm m}}\right)$. We
have
\begin{eqnarray*}
L_{F}^{\hbar}\left(H_{{\rm m}}\right) & = & \left[\omega^{ac}\partial_{c}\partial_{b}\frac{2}{\hbar}\left(\sum_{n=0}^{\infty}\frac{\left(-1\right)^{n}}{\left(2n+1\right)!}\left(\frac{\hbar}{2}\right)^{2n+1}\left(\omega^{ab}\partial_{b}^{\left(H\right)}\partial_{a}^{\left(\rho\right)}\right)^{2n}\right)H_{{\rm m}}c^{b}\frac{\partial}{\partial c^{a}}\right]\\
 & \equiv & \left[\frac{2}{\hbar}\widetilde{O}\omega^{ac}\partial_{c}\partial_{b}H_{{\rm m}}c^{b}\frac{\partial}{\partial c^{a}}\right]\,,
\end{eqnarray*}
where $\widetilde{O}$ abbreviates the sum in the round brackets of the first
line. We note that
\begin{eqnarray}
\omega^{ac}\partial_{c}\partial_{b}H_{{\rm m}}c^{b}\frac{\partial}{\partial c^{a}} & = & -\sum_{i\neq j}^{N}\xi^{i}\partial_{q_{i}}\partial_{q_{j}}H_{{\rm m}}\frac{\partial}{\partial\pi_{j}}\,.\label{eq:fermionic-Liouville-operator-xi-pi}
\end{eqnarray}
As far as the bosonic part $L_{B}^{\hbar}\left(H_{{\rm m}}\right)$
is concerned, we can follow the same logic as in section \ref{sub:Bosonic-BBGKY-hierarchy},
which applies also for the generalized density $\rho\left(\varphi^{a},c^{a}\right)$.
On the other hand, for $L_{F}^{\hbar}\left(H_{{\rm m}}\right)$, whenever
one of the indices $i,j$ in (\ref{eq:fermionic-Liouville-operator-xi-pi})
belongs to the $N-n$ particles set, we see that the differentiation with
respect to $\pi_{j}$ is actually a total derivative in the action
and therefore such contributions vanish. Thus effectively
\begin{eqnarray*}
\omega^{ac}\partial_{c}\partial_{b}H_{{\rm m}}c^{b}\frac{\partial}{\partial c^{a}} & = & -\sum_{i\neq j=1}^{n}\xi^{i}\partial_{q_{i}}\partial_{q_{j}}H_{{\rm m}}\frac{\partial}{\partial\pi_{j}}\,.
\end{eqnarray*}
By following exactly the same steps as in section \ref{sub:Bosonic-BBGKY-hierarchy}
and exploiting the symmetry under particle permutations we can rewrite
\begin{eqnarray*}
\omega^{ac}\partial_{c}\partial_{b}H_{{\rm m}}c^{b}\frac{\partial}{\partial c^{a}} & = & -\sum_{i\neq j=1}^{n}\xi^{i}\partial_{q_{i}}\partial_{q_{j}}\left(\sum_{k=1}^{n}\sum_{l=n+1}^{N}V\left(q_{k}-q_{l}\right)\right)\frac{\partial}{\partial\pi_{j}}\\
 & = & -\sum_{i\neq j=1}^{n}\xi^{i}\partial_{q_{i}}\partial_{q_{j}}\left(\left(N-n\right)\sum_{k=1}^{n}V\left(q_{k}-q_{n+1}\right)\right)\frac{\partial}{\partial\pi_{j}}\\
 & = & \left(N-n\right)\sum_{k=1}^{n}\omega^{ac}\partial_{c}\partial_{b}V\left(q_{k}-q_{n+1}\right)c^{b}\frac{\partial}{\partial c^{a}}\,.
\end{eqnarray*}
This implies that, in the action, we can integrate over the variables
$\chi_{n+2},\cdots\chi_{N}$ and conclude that also $L_{F}^{\hbar}$
acts on the reduced density $f_{n+1}$. Note that, on top of the terms
considered up to now, we also need to take into account the derivatives
present in $\widetilde{O}$. This does not modify the picture since
the very same consideration goes through applying the reasoning already
detailed in section \ref{sub:Bosonic-BBGKY-hierarchy}.

Our final result then reads:
\begin{eqnarray*}
\partial_{t}f_{n}+L_{B}^{\hbar}\left(H_{n}\right)f_{n}+L_{F}^{\hbar}\left(H_{n}\right)f_{n} & = & -\int d\chi_{n+1}\Biggr[\sum_{k=1}^{n}L_{B}^{\hbar}\left(V\left(q_{k}-q_{n+1}\right)\right)f_{n+1}\\
 &  & +L_{F}^{\hbar}\left(V\left(q_{k}-q_{n+1}\right)\right)f_{n+1}\Biggr]\,.
\end{eqnarray*}
This is our main result.
It reproduces equation (\ref{eq:bosonic-BBGKY-via-operator-L}) for zero form densities and
generalizes the standard BBGKY hierarchy by including higher order forms for the reduced densities
$f_{n}$.

The reader may wonder why one should consider an extended BBGKY hierarchy
that includes also higher density forms. The main reason is that the
ghosts are naturally present in the path integral formulation of out
of equilibrium statistical mechanics
\cite{Crossley:2015evo,Glorioso:2016gsa,Gao:2017bqf,Glorioso:2017fpd,Haehl:2015foa,Haehl:2015uoc,Haehl:2016pec,Haehl:2016uah,Haehl:2017zac,Jensen:2017kzi},
and as such one may wish to keep track of them also when considering
BBKGY. Moreover, as observed in the classical limit, i.e.~the CPI
reviewed in section \ref{sub:Classical-path-integral}, the ghosts have
proved to carry relevant physical information~\cite{Gozzi:1989xz,Gozzi:1993tm},
in particular in relation to the evolution of nearby trajectories (the so called Jacobi fields). 

Finally, further studies might lead to new insights. In fact let us
suppose that it is possible to supersymmetrize the Moyal formalism via
the recipe provided in~\cite{Gao:2017bqf}, then we would be lead to a field
theory on which we could exploit the powerful methods of supersymmetry.

\section{Summary and outlook} \label{sec:summary-outlook}

In this work we considered the functional formulation of the super-extended
Moyal formalism. Originally, this formalism was put forward to introduce
a proposal for differential calculus in quantum mechanics~\cite{Gozzi:1993nk}. It turns
out that this formalism is also related to the recently proposed
super-extension of the Schwinger-Keldysh formalism
\cite{Crossley:2015evo,Glorioso:2016gsa,Gao:2017bqf,Glorioso:2017fpd,Haehl:2015foa,Haehl:2015uoc,Haehl:2016pec,Haehl:2016uah,Haehl:2017zac,Jensen:2017kzi}.
In some sense,
by viewing the super-extended Moyal formalism from the point of view
of the Schwinger-Keldysh functional integral, one could say that there
are further arguments which lead to introduce the ghost sector of
the theory, in particular in relation to the unitarity condition (\ref{eq:unitarity-condition-SK}). 

In section \ref{sub:Comparison-with-the-SK-formalism} we considered
the topological properties of the path integral associated to the
Moyal formalism. In particular, we showed that, besides the topological
sector usually considered in the Schwinger-Keldysh formulation, one
can obtain a further topological sector by changing the boundary condition
of the path integral. Building on the CPI result~\cite{Gozzi:1989vv}, we showed
that the functional $Z_{\rm{pbc}}$ introduced in (\ref{eq:Zpbc-via-functional-integral-super_Moyal})
yields the Euler charachteristic of the phase-space manifold.

In section \ref{sec:BBGKY-hierarchy} we gave a functional derivation
of the quantum, super-extended BBGKY hierarchy. To the best of our
knowledge, a similar derivation was available only in the classical
limit~\cite{JolicoeurGuillou,Gozzi:1990pf,GozziReuterBBGKY}, where it proved useful to derive in a simple way
the Balescu-Lenard collision operator~\cite{JolicoeurGuillou}, see also~\cite{Chavanis1,Chavanis2}. 
The reason for considering
such an extended hierarchy is the following. As we said, in the functional
description of out of equilibrium phenomena a ghost sector naturally
arises in order to preserve the unitarity condition (\ref{eq:unitarity-condition-SK})
in a manifest way
\cite{Crossley:2015evo,Glorioso:2016gsa,Gao:2017bqf,Glorioso:2017fpd,Haehl:2015foa,Haehl:2015uoc,Haehl:2016pec,Haehl:2016uah,Haehl:2017zac,Jensen:2017kzi}. 
It is then natural
to keep track explicitly of this ghost sector and, possibly, a consistent
truncation of the BBGKY hierarchy has to contain also these further
ghost fields. On top of this, the ghost fields have the meaning of
Jacobi fields (i.e.~infinitesimal displacement of nearby trajectories)
in the classical limit and their evolution contains information regarding
the chaoticity of the system~\cite{Gozzi:1993tm}. 
Since in certain regimes the evolution of non equilibrium quantum fields can be well
approximated by classical systems~\cite{Aarts:2001yn} and 
the universal behaviour of far from equilibrium phenomena
can be sometimes described by nonthermal fixed points whose universality class encompass both 
quantum and classical systems~\cite{Orioli:2015dxa},
we believe that the introduction of these ghost fields carries
useful information. We hope to come back to these topics in the future.

\section*{Acknowledgments}

We would like to thank Ennio Gozzi and Martin Reuter for useful discussions.

\end{spacing}


\end{document}